\RequirePackage{fix-cm}
\documentclass[twocolumn]{svjour3}          % twocolumn
\smartqed  % flush right qed marks, e.g. at end of proof
\usepackage[T1]{fontenc}    % Determines font encoding of the output. Font packages have to be included before this line.
\usepackage[utf8]{inputenc} % Determines encoding of the input. All input files have to use UTF8 encoding.
\usepackage{graphicx}
\usepackage{wrapfig}
\usepackage{xurl,hyperref}

\usepackage{tabularx}
\usepackage{booktabs}
\usepackage{pifont}
\usepackage{ragged2e}
\usepackage{siunitx}
\usepackage{adjustbox}
\usepackage{array}
\usepackage{multirow,makecell}
\usepackage{inconsolata} % for nicer monospace in tables
\usepackage{dirtree} % for directory file trees
\usepackage{xcolor}
\usepackage[normalem]{ulem} % for \sout
\usepackage{listings}
\usepackage{minted}
\usepackage{todonotes}

\newcommand{\xmark}{\ding{55}} % ✗
\newcommand{\cmark}{\ding{51}} % ✓

\definecolor{bgcolor}{rgb}{0.95,0.95,0.92}
\definecolor{gray}{rgb}{0.4,0.4,0.4}
\definecolor{darkblue}{rgb}{0.0,0.0,0.6}
\definecolor{cyan}{rgb}{0.0,0.6,0.6}
\lstdefinelanguage{json}{
    basicstyle=\scriptsize\ttfamily,
    numbers=left,
    numberstyle=\tiny,
    breaklines=true,
    frame=single,
    backgroundcolor=\color{bgcolor},
    showstringspaces=false,
    string=[db]{"},
    captionpos=b,
    xleftmargin=1.5em,
    stringstyle=\color{green!50!black},
    morestring=[s][\color{blue!80}]{\ \ "}{":},
    keywordstyle=\color{cyan},
    morecomment = [s]{/*}{*/},
    commentstyle=\color{gray},
    keywords={true,false,null},
    moredelim=**[is][\color{blue!80!black}]{""}{":},
    literate=
     *{0}{{{\color{red}0}}}{1}
      {1}{{{\color{red}1}}}{1}
      {2}{{{\color{red}2}}}{1}
      {3}{{{\color{red}3}}}{1}
      {4}{{{\color{red}4}}}{1}
      {5}{{{\color{red}5}}}{1}
      {6}{{{\color{red}6}}}{1}
      {7}{{{\color{red}7}}}{1}
      {8}{{{\color{red}8}}}{1}
      {9}{{{\color{red}9}}}{1}
      {.}{{{\color{red}.}}}{1}
      {:}{{{\color{gray}{:}}}}{1}
      {,}{{{\color{gray}{,}}}}{1}
      {\{}{{{\color{gray}{\{}}}}{1}
      {\}}{{{\color{gray}{\}}}}}{1}
      {[}{{{\color{gray}{[}}}}{1}
      {]}{{{\color{gray}{]}}}}{1},
}

\newcolumntype{P}[1]{>{\centering\arraybackslash}p{#1}}
\newcolumntype{Y}{>{\centering\arraybackslash}X}
\usepackage{enumitem}

\newcounter{quotecount}

\newcommand{\ins}[1]{#1} % please insert
\newcommand{\del}[1]{} % please delete
\begin{document}

%\onecolumn
%\input{minor_revision/revision}
%\setcounter{section}{0}
%\setcounter{page}{1}
%\setcounter{table}{0}
%\setcounter{equation}{0}
%\twocolumn

\title{A Benchmarking Framework for Model Datasets%\thanks{Grants or other notes
%about the article that should go on the front page should be
%placed here. General acknowledgments should be placed at the end of the article.}
}
%\subtitle{Do you have a subtitle?\\ If so, write it here}

\titlerunning{A Benchmarking Framework for Model Datasets}        % if too long for running head

\author{Philipp-Lorenz Glaser \and 
        Lola Burgueño \and
        Dominik Bork
}

%\authorrunning{Short form of author list} % if too long for running head

\institute{Philipp-Lorenz Glaser \at
              TU Wien, Business Informatics Group, Vienna, Austria \\
              \email{philipp-lorenz.glaser@tuwien.ac.at}
              \and
              Lola Burgueño \at
              ITIS Software, University of Malaga \\
              \email{lolaburgueno@uma.es}           %  \\
%             \emph{Present address:} of F. Author  %  if needed
           \and
           Dominik Bork \at
              TU Wien, Business Informatics Group, Vienna, Austria \\
              \email{dominik.bork@tuwien.ac.at}
}

\date{Received: date / Accepted: date}
% The correct dates will be entered by the editor

\maketitle

\begin{abstract}
Empirical and LLM-based research in model-driven engineering increasingly relies on datasets of software models, for instance, to train or evaluate machine learning techniques for modeling support. These datasets have a significant impact on solution performance; hence, they should be treated and assessed as first-class artifacts. However, such datasets are typically collected or created ad hoc and without guarantees of their quality for the specific task for which they are used. This limits the comparability of results between studies, obscures dataset quality and representativeness, and leads to weak reproducibility and potential bias.
In this work, we propose a benchmarking framework for model datasets (i.e., benchmarking the dataset itself). Benchmarking datasets involves systematically measuring their quality, representativeness, and suitability for specific tasks. 
To this end, we propose a Benchmark Platform for MDE that provides a unified infrastructure to systematically assess and compare software model datasets across modeling languages and formats, using defined criteria and metrics.

  % \PACS{PACS code1 \and PACS code2 \and more}
  % \subclass{MSC code1 \and MSC code2 \and more}
\end{abstract}

\section{Introduction}\label{intro}

Model-driven engineering (MDE) is a well-established field that emphasizes the use of high-level models as primary artifacts throughout the software engineering process. Now, with the rapid rise of AI, MDE is entering a new phase in which intelligent automation can augment traditional model creation, transformation, analysis, and more. Together, MDE and AI offer promising prospects for mutual enrichment and collaboration. 
Many initiatives have arisen over the years to apply AI in advancing MDE, such as theme sections in journals that \del{received a good number of papers}\ins{attracted multiple contributions}~\cite{BurguenoCWZ22,BurguenoDRB26} and the international workshop series on Artificial Intelligence and Model-Driven Engineering (MDE Intelligence)\footnote{\url{https://mde-intelligence.github.io/}}, which was started at MODELS 2019~\cite{BurguenoMDEIntelligence19} and, since then, has emerged from a specialized small workshop into a vibrant core workshop of MODELS, attracting the most submissions among all workshops.

However, for the integration of AI and MDE to materialize and flourish, as well as to enable systematic and structured progress as a scientific community, standardized methods and tools for the analysis and comparison of research datasets are essential. In the context of MDE, these datasets take the form of collections of models\footnote{Note that in this work we refer to `model' and `conceptual model' as models conforming to modeling languages UML, ArchiMate, and BPMN, whereas we refer to machine learning models or ML models when referring to trained AI models.} (such as UML models) used to train AI systems. Establishing consistent standards for their evaluation and comparison is therefore crucial to advancing research in this area~\cite{CamaraBT24}.

Available model datasets often have quality issues such as clones and dummy models~\cite{Djelic2025pipeline,storrle2010towards}. Recently, Burgue\~no et al. even stated that the ``scarcity of high-quality datasets hinders the advancement of AI-driven automation in MDE and benchmarks for training and evaluating AI models.''~\cite{Burgueno25AutomationMDE}

%\todo{expand introduction}

This situation highlights a growing need for consolidated efforts within the MDE community to better understand the characteristics and limitations of existing model datasets. As AI techniques often depend heavily on data quality, even subtle flaws in model repositories can propagate through multiple stages of the engineering process, leading to misleading conclusions or biased automation behaviors.
%Without deeper insights into the datasets that are used as part of Machine Learning tasks, it becomes difficult to assess the reliability of AI-assisted MDE outcomes.
The absence of such an understanding not only hinders reproducibility but also complicates the interpretation of experimental results, as different studies may rely on fundamentally incomparable datasets.
This reinforces the importance of sound data management practices and a shared awareness of dataset quality. Developing a common ground for how datasets should be analyzed and compared is needed. %more needed than ever before.

ML and LLM-based tasks may depend strongly on naming quality and lexical richness, or on structural variety and complexity for classification, completion, repair, and refactoring tasks. Not all of these characteristics can be easily and uniformly benchmarked. The role of a benchmarking framework is therefore not to impose a single notion of quality, but to make salient characteristics explicit and, where possible, measurable, so that researchers can judge task suitability, report dataset properties transparently, and improve comparability and reproducibility across studies. As Di Rocco et al. stated, ``To gauge the effectiveness of LLMs integration in MDE tasks, establishing standardized benchmarks becomes crucial.''~\cite{di2025use}

\ins{Our focus is dataset characterization rather than dataset cleansing. The framework computes reproducible evidence about the models contained in a dataset under an explicit benchmark profile, but it does not aim to transform the input into a cleansed replacement dataset. Cleansing decisions, which may change the underlying dataset, such as removing dummy models, filtering languages, or eliminating duplicate and near-duplicate models are modeling language- and task-dependent, and therefore not trivial. We treat benchmarking as an evi\-dence-producing step that can inform subsequent curation or cleansing steps (e.g., MCP4CM~\cite{Djelic2025pipeline}), but remains conceptually separate from them.}

In this paper, we propose a benchmarking framework and a platform that implements it. This platform accepts models expressed in languages such as UML, ArchiMate, and Ecore, parses them into a normalized intermediate representation, and extracts descriptive and structural statistics through a dedicated metrics engine. The results are compiled into benchmark reports that summarize dataset characteristics, support cross-dataset comparisons, and enable researchers to reason about datasets.
We outline a research agenda to (1) study existing benchmark approaches from related domains (e.g., software engineering or machine learning), (2) formalize a core set of metrics, (3) develop a prototype implementation for selected modeling languages and formats, and (4) evaluate its utility on existing model datasets and published research in the MDE community.

The remainder of this paper is structured as follows. Section~\ref{sec:background} reviews the related work. Section~\ref{sec:cm-datasets} presents the background on model datasets. Section~\ref{sec:framework} introduces a benchmarking framework to enable systematic evaluation, while Section~\ref{sec:prototype} presents the corresponding platform implementation. Section~\ref{sec:demonstration} demonstrates the use of the platform on three representative datasets. Finally, Section~\ref{sec:discussion} discusses the key findings, lessons learned, limitations, and future directions, and Section~\ref{sec:conclusion} concludes the paper.

\section{Related Work}\label{sec:background}

Model-Driven Engineering (MDE) relies on benchmarks to evaluate tools, transformations, and scalability in handling complex models. 
Benchmarks provide a means to compare approaches on common grounds, assess performance under controlled conditions, and ensure reproducibility of results across studies. In the following, we briefly review relevant work on benchmarking within the Software Engineering and Model-driven Engineering communities.

\paragraph{Benchmarks in Model-driven Engineering:}

Benelallam et al.~\cite{BenelallamTRIK14} gathered and made publicly available a set of four benchmarks for model transformation and query engines, derived from real-world MDE scenarios with models of increasing size, either concrete (e.g., reverse-engineered Java projects) or generated via customizable deterministic instantiators.

Str\"uber at al. are concerned with scalability challenges of growing model transformation rule sets themselves and address maintainability and performance issues in large MDE specifications by introducing three benchmarks with expanding rule sets as a shared community resource for standardized evaluation of transformation engines~\cite{Strueber16}.

Sauerwein~\cite{sauerwein2013} proposed a model-driven approach for automatically generating benchmark data with their corresponding ground truth. 

Zhu et al.~\cite{Zhu06} propose a Model-Driven Architecture-based approach to automatically generate customized performance benchmark suites for Web services from UML architectural and testing models.

\ins{Tolvanen et al.~\cite{TolvanenKRPT25} introduce a framework for evaluating tool support for the co-evolution of modeling languages and models. Their framework defines language-change scenarios (e.g., add, rename, remove) over abstract syntax (i.e., metamodel), concrete syntax (i.e., notation), and constraints, and then scores the impact of these changes. It is similar to our work in that both approaches provide a structured, repeatable evaluation framework for MDE artifacts and tooling. However, the evaluation target is different, as they assess the co-evolution capabilities of language workbenches such as MetaEdit+, EMF/Sirius, and Jjodel, while our framework assesses the technical usability and descriptive characteristics of model datasets independent of a particular modeling workbench or co-evolution scenario.}

\ins{Derks et al.~\cite{DerksSB23} propose vpbench, a framework for generating version histories to benchmark techniques for evolving variant-rich software systems. Their framework simulates feature-oriented evolution histories (e.g., adding, removing, cloning, and transplanting features) and records metadata, such as feature models, feature locations, clone traces, and developer-intention metadata, as ground truth. The work is related to ours in its goal of improving benchmark maturity, reproducibility, and transparency through explicit metadata and an extensible framework. However, its benchmark subjects are generated software version histories used to evaluate evolution techniques, whereas our framework characterizes existing model datasets.}

\paragraph{Benchmarks in Artificial Intelligence and Software Engineering:}
\cite{koohestani2025} presents a systematic review of 273 AI for software engineering benchmarks from 247 studies between 2014 and 2025. It also introduces a semantic search tool for efficient discovery of benchmarks.

\paragraph{Benchmarks in Artificial Intelligence and Model-driven Engineering:}
ModelXGlue~\cite{lopez2025modelxglue} proposes a benchmarking framework for ML tools in MDE. It allows researchers to build benchmarks for tasks like model classification, clustering, and feature recommendation.

Camara et al.~\cite{CamaraBT24} present a conceptual framework to standardize LLMs benchmarking in software modeling, addressing variability in prompts, outputs, and solution spaces.

\paragraph{Synopsis:}
Although all of these works propose benchmarking software artifacts or generating datasets for benchmarking, none address benchmarking datasets in the context of Model-Driven Engineering.

% \subsection{Synopsis}

%%BEGIN COPY FROM MODELS25 PAPER
% Numerous open Modelsets have been proposed in various modeling languages such as ArchiMate~\cite{Glaser25-EAModelset} (977 models), Petri Nets~\cite{hillah2017petri} (664 models), BPMN~\cite{corradini2019reprository} (174 models) \cite{saeedi2025empirical} (25,866 models) \cite{compagnucci2021trends} (25,590 models), Ecore~\cite{onder_babur_2019_2585456} (555 models), OntoUML~\cite{barcelos2022} (185 models) and UML~\cite{LopezIC22} (10,586 models)~\cite{robles2017extensive} (93,000 models). From these, only~\cite{LopezIC22},~\cite{saeedi2025empirical} and~\cite{onder_babur_2019_2585456} provide a labeled model dataset where each model is tagged with relevant tags, e.g. for the domain.
%%END COPY FROM MODELS25 PAPER

%GoldenUML Modelset~\cite{Verbruggen25.GoldenModelset}

\section{Background: Model Datasets}\label{sec:cm-datasets}
%With the increasing interest in data-driven and ML-enhanced MDE, several model datasets have been proposed. 
Model datasets differ from conventional corpora such as text, image, audio, or source-code collections in that their primary objects are \del{formal models} \ins{modeling artifacts} whose content is defined by a modeling language and whose structure is constrained by a metamodel. As a result, model datasets are not merely collections of semi-structured text files but collections of typed, interrelated artifacts with explicit abstract syntax, well-formedness constraints, and varying tool-dependent representations.

Notably, existing model datasets often do not report comprehensive descriptive properties, e.g., language coverage, size, curation, and intended use. To reason about them and to interpret benchmark results, it is necessary to understand both the landscape of existing datasets and the intrinsic properties that characterize them. Therefore, this section \textit{(i)} abstracts from these examples to identify characteristic properties of model datasets (\autoref{sec:cm-dataset-characteristics}), \textit{(ii)} surveys representative model datasets used in current research (\autoref{sec:existing-datasets}), and \textit{(iii)} discusses a typical lifecycle of model dataset construction and use (\autoref{sec:cm-dataset-lifecycle}), highlighting where benchmarking can provide actionable evidence.

\subsection{Characteristics of Model Datasets}\label{sec:cm-dataset-characteristics}
% Intro
We now provide the conceptual grounding for the remainder of the paper by clarifying what constitutes a model dataset and which intrinsic properties are particularly relevant for reasoning about the underlying data. It also delineates the scope boundaries because not all practically relevant aspects of datasets can be assessed uniformly.

% What is a CM Dataset?
In this work, a model dataset is a curated collection of model artifacts expressed in one or more modeling languages (e.g., UML, BPMN, ArchiMate, Ecore). Models are typically stored in machine-readable formats such as XMI, XML, or JSON, or in textual notations such as PlantUML. Models can also be available only as rendered diagrams (for example, images embedded in documents or exported as PNG), but such representations require fundamentally different processing techniques (e.g., image segmentation, object detection, symbol recognition) and are therefore out of scope for our current benchmarking approach. A model may be stored in a single file or distributed across multiple files, either due to explicit modularization or due to format-specific persistence strategies (e.g., ArchiMate models can be stored using the GRAFICO format\footnote{https://github.com/archi-contribs/archi-grafico-plugin/wiki/GRAFICO-explained}, which serializes model elements into separate XML files). Models may also reference external resources, such as imported libraries or other model elements that are required for correct interpretation. Some datasets, in addition to models, provide other artifact types per case, such as textual requirements, source code, images, or reports. While these artifacts can be important for specific research questions, they are highly heterogeneous and rarely share a common structure across datasets. For this reason, we do not deal with non-model artifacts in this work. % as out of scope in our assessments.

% Granularity
Model datasets exhibit a clear granularity hierarchy. At the top level, \textit{dataset-level properties} emerge from the set of contained models and their distributions. At the next level, \textit{model-level properties} reflect the internal composition of individual models. At the lowest level, \textit{element-level properties} capture characteristics of individual model elements and relationships. Thus, within a model, we distinguish between the model artifact as a whole and the model elements and relations it contains. Model elements are instances of the language's abstract syntax (e.g., UML Class, Ecore EClass). Many modeling languages additionally support multiple diagrammatic views on the same underlying abstract syntax (e.g., overlap in UML class and sequence diagrams, ArchiMate views), which further contributes to the layered nature of model datasets.

% Metadata & Annotation
Next to the models themselves, model datasets may contain \textit{metadata} and \textit{annotations}. Metadata can appear at dataset-, model-, or even element-level and includes provenance information, authorship, timestamps, tool identifiers, version information, licensing statements, and documentation. Annotations are often task-oriented and serve as ground truth for downstream tasks (e.g., clustering, classification), for example, domain labels, quality flags, and defect markers. High-quality annotations are often expensive because they require expert judgment, particularly for semantic properties such as domain correctness or modeling quality, and are therefore still relatively rare. Although metadata and annotations are crucial for interpretation, reproducibility, and task design, their representations vary widely across datasets and are rarely comparable without task-specific assumptions. In this work, we therefore treat annotations and metadata as contextual enrichment and focus our initial benchmarking framework on characteristics that are intrinsic to the model artifacts.

% Formats
The model representation \textit{format} in model datasets is a major source of heterogeneity and a primary driver of benchmarking difficulty. With respect to modeling languages, datasets may be single-language (e.g., only ArchiMate models) or multi-language (e.g., mixing UML and Ecore), and even within a single language, different metamodel versions, tool-specific dialects, and conventions are common. Serializations and tooling introduce further heterogeneity, as each dataset stores models in different formats depending on their origin and subsequent processing. Furthermore, different formats preserve different information. Some serializations include diagram layout and styling (coordinates, colors, fonts), others encode only the abstract syntax, and some omit, flatten, or approximate specific constructs and features. As a result, seemingly homogeneous datasets may exhibit variation in parseability and completeness, and the same model can yield different internal representations depending on the parser and serialization.

% Size & Composition
\textit{Size} and composition further contribute to heterogeneity. Typical model datasets contain hundreds to a few thousand models, rather than millions, and individual models range from small sketches with a handful of elements to large industrial artifacts with hundreds or thousands of elements and relationships. Thus, size distributions are often strongly skewed, with many ``toy'' or ephemeral models and a small number of very large, long-lived models. Construct usage also tends to be imbalanced, as certain metaclasses with specific features often dominate, while rare language features may be absent or underrepresented, leading to construct sparsity that can bias evaluations for tasks that rely on specific constructs. Outliers are common and include unusually large or unusually small models, atypical modeling practices (for example, using ArchiMate primarily as a class-diagram-like notation), or extreme naming conventions that inflate counts without adding meaningful variability. Duplicates and near-duplicates are also frequent as models may appear in multiple slightly modified versions, for example, due to teaching material variants, refactorings, forks, or repeated reuse of case-study models. Such redundancy can distort descriptive statistics, inflate performance in supervised learning, and violate independence assumptions in evaluation, yet it is non-trivial to detect reliably. Finally, the mix of application domains represented in a dataset varies widely and strongly affects the generalizability of empirical findings, for instance, when conclusions are drawn from models drawn almost exclusively from a single industry sector.

% Acquisition & Origin
The way models in model datasets are \textit{acquired} and curated further shapes their properties. Datasets may be manually curated by researchers or educators, mined from model repositories or version-control platforms (e.g., GitHub)~\cite{dyer2015boa}, or generated synthetically, e.g., through transformations, mutation~\cite{gomez2018tool}, or LLMs~\cite{conrardy2024image,LLM-Model-Instantiation.26}. \textit{Origins} matter because they often correlate with noise patterns. Educational models often contain relatively small models that represent artificial or simplified scenarios, but they are usually clean (unless exercise variations). Mined datasets tend to be noisier~\cite{saeedi2025empirical} and include many sketches and partial models, but they may better reflect everyday modeling practice. Industrial models are typically larger and validated through real-world use, yet they are rarer and often constrained by confidentiality. Curation practices such as inclusion criteria, deduplication policies, normalization steps, and provenance tracking influence bias and must be understood to draw conclusions about the underlying data.

% Non-technical constraints & Task-specificity
Finally, several non-technical constraints are crucial for practical use of the dataset but are only partially amenable to automated benchmarking. \textit{Licensing and redistribution} conditions determine whether a dataset can be shared and reused. \textit{Provenance and attribution} affect interpretability and reproducibility. \textit{Confidentiality}, \textit{anonymization}, and \textit{ethical} considerations can arise, e.g., for industrial models or sensitive domains, and may restrict what can be released or how it can be processed. 
From a task perspective, different \textit{intended uses} emphasize distinct characteristics, and there is an urgent need for a benchmarking framework to make salient characteristics explicit and, where possible, measurable, so that researchers can assess task suitability, report dataset properties transparently, and improve comparability and reproducibility across studies.

To yield a standardized, modeling-language-agnostic representation format for models, we treat models as (and transform them into) structured, typed graphs constrained by a metamodel, rather than as untyped token sequences. Models typically combine containment hierarchies (e.g., attributes contained in classes, which are further contained in packages) with cross-references through identifiers (i.e., elements linked through source and target identifiers in relationships). Metamodel conformance shapes virtually all structural properties, e.g., through typing rules, multiplicities, or containment/reference constraints. This yields high semantic density as small modifications, such as changing a type, a multiplicity, or the endpoint of a relationship, can imply substantial semantic changes in the modeled system. Element names and labels carry additional important information, but they represent only one layer alongside structural typing, relationships, and constraints. For benchmarking, this implies that purely lexical statistics are insufficient and must be complemented by structural measures that reflect metamodel-driven properties.

\subsection{Existing Model Datasets}\label{sec:existing-datasets}

\begin{table*}[t]
\centering
\caption{Classification of selected model datasets.}
\label{tab:cm-datasets}
\scriptsize
\setlength{\tabcolsep}{2.2pt}
\renewcommand{\arraystretch}{1.08}

\begin{tabularx}{\textwidth}{@{}
p{2.2cm}
p{1.6cm}
X
p{2.05cm}
p{2.25cm}
p{2.2cm}
X
X
@{}}
\toprule
\textbf{Dataset} &
\makecell[l]{\textbf{Modeling}\\\textbf{language}} &
\textbf{Format} &
\textbf{Acquisition} &
\textbf{Origin} &
\makecell[l]{\textbf{Size}\\\textbf{(\#Models)}} &
\textbf{Metadata} &
\textbf{Annotations} \\
\midrule

ModelSet~\cite{LopezIC22} &
UML, Ecore &
\RaggedRight UML XMI + Ecore XMI\arraybackslash &
\RaggedRight Mined + curated (light cleaning)\arraybackslash &
\RaggedRight GenMyModel (UML), GitHub (Ecore)\arraybackslash &
\RaggedRight 10{,}586 (5{,}466 Ecore, 5{,}120 UML)\arraybackslash &
\RaggedRight \cmark \ (relational DB with model-level metadata)\arraybackslash &
\RaggedRight \cmark \ category, tags, purpose, notation, tool, etc.\ as labels\arraybackslash \\

\midrule

EA ModelSet~\cite{Glaser25-EAModelset} &
ArchiMate &
\RaggedRight JSON + XML (Open Exchange / Archi) + CSV\arraybackslash &
\RaggedRight Mined + curated (light cleaning, normalization)\arraybackslash &
\RaggedRight GitHub, GenMyModel, community/other sources\arraybackslash &
\RaggedRight 977\arraybackslash &
\RaggedRight \cmark \ (FAIR-based metadata)\arraybackslash &
\RaggedRight \xmark\arraybackslash \\

\midrule

Golden UML Dataset~\cite{Verbruggen25.GoldenModelset} &
\RaggedRight UML (class diagrams)\arraybackslash &
\RaggedRight PlantUML + Markdown description + PNG/SVG images\arraybackslash &
\RaggedRight Curated (manually selected teaching cases)\arraybackslash &
\RaggedRight Education\arraybackslash &
\RaggedRight 45\arraybackslash &
\RaggedRight \cmark \ (per-case metadata files)\arraybackslash &
\RaggedRight \cmark \ (domain labels / tags in metadata)\arraybackslash \\

\midrule

SAP-SAM Signavio Dataset~\cite{sola2022signavio} &
\RaggedRight BPMN, Value Chain, DMN, CMMN, EPC, UML, ArchiMate, \ldots\arraybackslash &
\RaggedRight JSON (SAP Signavio format)\arraybackslash &
\RaggedRight Mined from platform + lightly curated/anonymized\arraybackslash &
\RaggedRight Signavio Academic Platform\arraybackslash &
\RaggedRight 1{,}021{,}471 (618{,}807 BPMN, 194{,}078 Value Chain, 10{,}956 ArchiMate, etc.)\arraybackslash &
\RaggedRight \cmark \ (platform-derived properties)\arraybackslash &
\RaggedRight \xmark\arraybackslash \\

\midrule

Lindholmen Dataset~\cite{robles2017extensive} &
UML &
\RaggedRight Links + mixed file types (UML XMI, diagram images)\arraybackslash &
\RaggedRight Mined\arraybackslash &
\RaggedRight GitHub\arraybackslash &
\RaggedRight $\approx$93{,}000\arraybackslash &
\RaggedRight \cmark \ (SQL + CSV with file-level metadata)\arraybackslash &
\RaggedRight \xmark\arraybackslash \\

\midrule

AtlanMod Zoo\footnote{\url{https://github.com/atlanmod/atlantic-zoo/tree/main}\label{foot-atlanmod}} &
\RaggedRight Ecore, UML, \ldots\arraybackslash &
\RaggedRight Ecore XMI + mixed formats (e.g., UML XMI, OWL, GraphML, custom DSL, \ldots)\arraybackslash &
\RaggedRight Curated\arraybackslash &
\RaggedRight AtlanMod/related tooling projects\arraybackslash &
\RaggedRight $\approx$300\arraybackslash &
\RaggedRight \xmark\arraybackslash &
\RaggedRight \xmark\arraybackslash \\

\midrule

OntoUML/UFO Catalogue~\cite{barcelos2022} &
\RaggedRight OntoUML (+ UFO grounding)\arraybackslash &
\RaggedRight JSON + Turtle + Visual Paradigm (.vpp) + PNG images\arraybackslash &
\RaggedRight Curated\arraybackslash &
\RaggedRight Academic and industrial case studies from literature, teaching, and projects\arraybackslash &
\RaggedRight 127\arraybackslash &
\RaggedRight \cmark \ (FAIR-based, rich metadata)\arraybackslash &
\RaggedRight \xmark \arraybackslash \\

\midrule

Labeled Ecore Dataset~\cite{onder_babur_2019_2585456} &
Ecore &
\RaggedRight Ecore XMI\arraybackslash &
\RaggedRight Mined\arraybackslash &
\RaggedRight GitHub\arraybackslash &
\RaggedRight 555\arraybackslash &
\RaggedRight \xmark\arraybackslash &
\RaggedRight \cmark \ (domain labels)\arraybackslash \\

\bottomrule
\end{tabularx}
\end{table*}

In the following, we survey existing model datasets and relate them to some of the characteristics discussed previously. \autoref{tab:cm-datasets} provides an overview of the main properties of each dataset, including modeling languages, formats, acquisition strategy, origin, size, and the presence of metadata and annotations.

\ins{
Since many modeling languages and notations exist, we focus in this work on datasets that contain models in ArchiMate, Ecore, and UML. This focus is a scoping decision for the initial survey and prototype, as reusable datasets for these widely used modeling languages are comparatively visible and accessible in the MDE community. However, our framework is not conceptually limited to such structural, general-purpose languages, and in the future, we are interested in extending the framework to support domain-specific, executable, and simulation-oriented modeling languages, e.g., BPMN~\cite{corradini2019reprository,saeedi2025empirical}, Simulink~\cite{BollBAKV21}, or Petri Nets~\cite{hillah2017petri}.
}

\del{
Since many modeling languages and notation exist, in this work we focus on datasets that contain models in ArchiMate, Ecore, or UML.
%as our implemented benachmarking platform (see \autoref{sec:prototype}) currently fully supports these three languages.
Although our work is extensible to any modeling language, model datasets of models conforming to other modeling languages, e.g., BPMN~\cite{corradini2019reprository,saeedi2025empirical,compagnucci2021trends} or Petri Nets~\cite{hillah2017petri} are left for future work.
% exist and we plan to extend our platform to support also these languages. %In the following, we describe each dataset in more detail.
% \todo{Correct references for the BPMN datasets from MODELS25 paper}
}

ModelSet~\cite{LopezIC22} combines models mined from GenMyModel (UML) and GitHub (Ecore) with light curation and cleaning, resulting in a labeled corpus of 10,586 models (5,120 UML and 5,466 Ecore). Beyond the model files (UML XMI and Ecore XMI), ModelSet provides a relational database with model-level metadata and a comparatively rich set of labels (e.g., category, tags, purpose, notation, tool). Its size and diversity make it one of the few resources suitable for supervised ML on software models.

EA ModelSet~\cite{Glaser25-EAModelset} is a FAIR\footnote{FAIR data refers to data that meets international guidelines for research data management, ensuring it is Findable, Accessible, Interoperable, and Reusable for both humans and machines~\cite{wilkinson2016fair}.} dataset of ArchiMate models and comprises 977 models collected from GitHub, GenMyModel, and other community sources. Models are stored in JSON, XML (Open Exchange format and Archi storage format), and CSV. In contrast to ModelSet, it is metadata-rich (per FAIR principles) rather than annotation-rich, as it does not yet include explicit task-specific annotations.

The Golden UML Dataset~\cite{Verbruggen25.GoldenModelset} is a semantically rich collection of 45 UML class-diagram cases drawn from educational settings. Each case couples a natural language description (in Markdown) with a PlantUML class diagram specification and corresponding diagram images (PNG/SVG), plus per-case metadata files (with domain labels and tags). Compared to mined corpora, its value lies in interpretability and high-quality ``ground truth'' modeling intent rather than large-scale statistical analysis.

The SAP-SAM Signavio Dataset~\cite{sola2022signavio} exemplifies the opposite extreme in terms of scale. It consists of over one million models created on the Signavio Academic Platform over a ten-year period, with BPMN and value chain models dominating, but multiple other notations are present, including DMN, CMMN, EPC, UML, and ArchiMate. Models are stored in the platform's JSON format and derived from real usage by thousands of users (who are anonymized in the dataset). It includes platform-derived properties as metadata (e.g., user information, timestamps), but no explicit annotations. It thus offers a highly realistic, noisy, multi-notation corpus for large-scale empirical analysis.

The Lindholmen Dataset~\cite{robles2017extensive} is another large, mined corpus focused on UML artifacts in open-source projects. It contains roughly 93,000 UML-related files discovered on GitHub, including mostly diagram images and a few thousand UML XMI files. The acquisition process is fully automated, and the dataset is accompanied by SQL and CSV files with file-level metadata (e.g., repository URLs). There are no annotations.

AtlanMod Zoo is a long-standing curated collection of metamodels and related artifacts originating from the AtlanMod group and associated tooling projects. It contains about 300 Ecore XMI models, which are also available in other formats (e.g., UML XMI, OWL, GraphML, etc.). It provides no metadata and annotations; instead, it serves as a diverse set of metamodels, which have been widely reused in MDE research (e.g.,~\cite{BaburC17,YangS22}).

The FAIR OntoUML/UFO Catalog~\cite{barcelos2022} is a curated collection of 127 OntoUML models grounded in the UFO foundational ontology, drawn from the OntoUML community. Models are provided in multiple formats, including JSON, RDF/Turtle, Visual Paradigm projects, and diagram images, and are published with rich metadata. The catalog does not include any explicit task-specific annotations.

Finally, the Labeled Ecore Dataset~\cite{onder_babur_2019_2585456} provides 555 Ecore metamodels mined from GitHub and manually assigned to application domains. Models are stored as Ecore XMI files without additional structured metadata, but with explicit domain labels used for clustering and classification experiments.

\subsection{Lifecycle of Model Datasets}\label{sec:cm-dataset-lifecycle}
\begin{figure}
    \centering
    \includegraphics[width=1\linewidth]{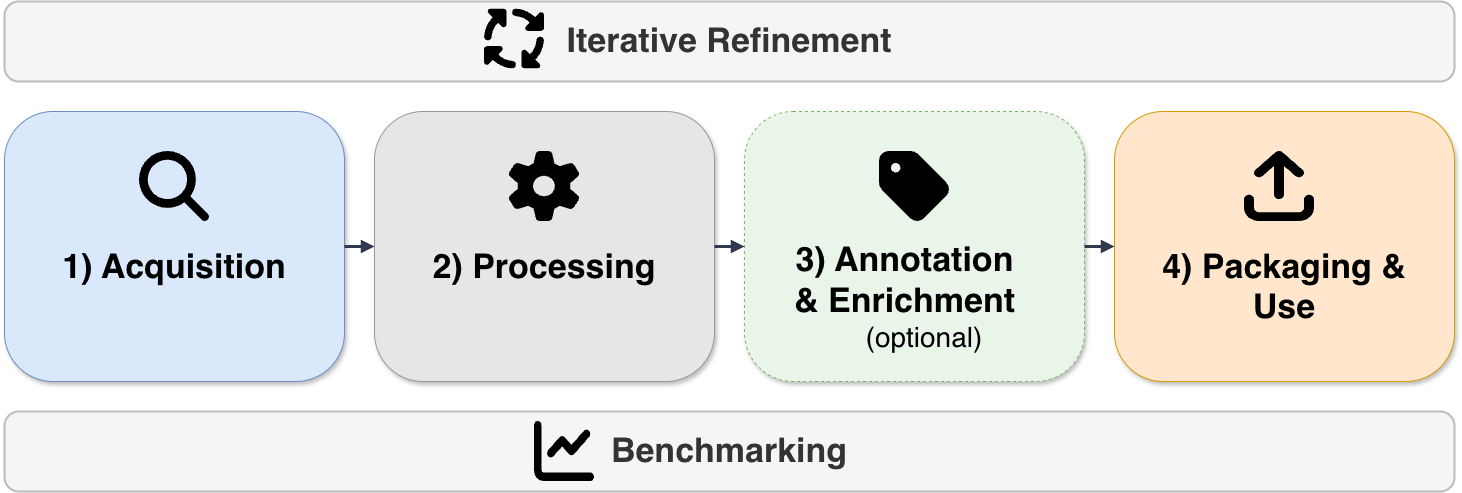}
    \caption{Lifecycle of Model Datasets}
    \label{fig:lifecycle}
\end{figure}

Model datasets typically emerge through a sequence of collection and transformation steps rather than as ready-to-use research assets. A lifecycle (cf. \autoref{fig:lifecycle}) perspective is useful because many dataset properties that later affect comparability and validity are the result of earlier choices, for example, where models were obtained, which artifacts were included, how parsing failures were handled, or which variants were filtered. While the phases discussed below are conceptually ordered, model dataset work is rarely linear. It is usually iterative and overlapping, with observations from later stages triggering revisions to earlier steps, such as adjusting acquisition criteria, normalization rules, or annotations.

\subsubsection{Acquisition}
In the \textbf{acquisition} phase, models are added to the dataset, e.g., by collecting them from existing sources or generating them synthetically. Typical sources include mining open-source model repositories (e.g., GenMyModel), version control platforms (e.g., GitHub)~\cite{dyer2015boa}, community-curated collections, teaching materials, industrial collaborations, and previously published datasets. Acquisition often involves applying file-type filters, handling tool-specific export formats, and capturing basic provenance when available, such as source location, tool version, timestamps, and licensing information. In addition to mining, models may be generated synthetically, for example, via generators, model mutation~\cite{gomez2018tool}, or LLMs that produce models from prompts~\cite{conrardy2024image}. 

In practice, hybrid datasets are common, combining, for example, mined artifacts with manually curated examples or synthetic additions. These acquisition decisions strongly shape later benchmarking outcomes, because they determine the raw heterogeneity of languages and formats, the completeness of referenced dependencies, and the representativeness of domains and modeling styles. The result of acquisition is therefore best understood as a raw corpus of model artifacts with optional preliminary provenance, but with limited guarantees of consistency, parseability, or quality. Even at this early stage, benchmarking can provide triage evidence, for instance, by quantifying the mix of languages and formats, the proportion of files likely to be parseable, or the rough size distribution of artifacts, which inform subsequent processing decisions.

\subsubsection{Processing}\label{sec:cm-dataset-processing}

\ins{The second phase is \textbf{processing}, which converts a raw corpus into a technically usable, more comparable dataset. This phase is particularly model-specific as it requires metamodel-aware handling of structure, constraints, and dependencies. Processing typically begins with parsing and importing artifacts into a controlled toolchain or an intermediate representation (e.g., a graph-based representation or a language-specific object graph).}

\ins{The concrete meaning of parsing depends on the modeling language, the serialization format, and the intended analyses. Many models are stored as serializations of a metamodel-based language, e.g., XMI documents for models conforming to an EMF-based metamodel. If the relevant metamodel and libraries are available, frameworks such as EMF (Java) or PyEcore (Python) can be used to load the serialized artifact into an object graph and expose typed elements, attributes, containment relations, and references. This strategy can support metamodel-aware processing and validation, but it also requires careful handling of metamodel versions, namespace changes, URI/path mappings, primitive libraries, and external resources, especially when a dataset contains models exported by different tools or language versions. Alternatively, parsers may operate directly on the serialization, e.g., by reading XML, JSON, or a custom textual format and applying extraction rules for the relevant constructs. Such serialization-oriented parsing is useful when no suitable metamodel implementation is available, when only a reduced view of the model is needed (e.g., names, labels, element types, or selected relations), or when the source format is tool-specific. The trade-off is that the parser must explicitly define which concepts and references are reconstructed, and analyses are then limited to this implemented coverage.}

\ins{The parsing step, therefore, determines which subsequent analyses are feasible. Checks over, e.g., element types, labels, or other properties, can only be performed when the parser has loaded or reconstructed the corresponding information. Processing thus involves handling multiple formats and versions, resolving cross-references and external resources where possible, and recording parsing outcomes, warnings, skipped elements, and failures. Validation likewise depends on the selected parser and the available language support. It may include metamodel conformance checks, language-specific well-formedness constraints, duplicate identifiers, constraint violations, or dangling references, but only when the processing toolchain represents the information needed for these checks.}

\ins{After parsing and validation, normalization and canonicalization reduce non-semantic variability that would otherwise impact statistics and comparisons, e.g., standardizing identifiers, aligning different language versions, or harmonizing types and names. Depending on the intended uses, processing may also separate semantic content from incidental data, such as layout or tool metadata, when they are not relevant to the intended tasks. Models that are corrupted, trivially small, unsupported by the selected parser, or otherwise unusable may be removed, quarantined, or explicitly tagged. A further challenge is handling redundancy, such as detecting duplicates and near-duplicates, identifying model families (variants, forks), and deciding whether to collapse them, keep them with explicit tags, or filter them.}

\ins{The outcome of this stage is a processable dataset with known parsing status, normalized representations, and explicit decisions about inclusion and redundancy. Benchmarking is particularly informative at this stage because it can quantify what changed relative to the raw corpus, for example, improved parseability, reduced variance due to normalization, shifts in size and construct distributions induced by filtering, or reductions in duplication that improve evaluation reliability.}

\subsubsection{Annotation and Enrichment}\label{sec:annotation-enrichment}
\del{A third optional phase is \textbf{annotation and enrichment}, which introduces task-specific information layers on top of a base dataset. Annotations range from coarse model-level labels such as domain categories to fine-grained markings of elements, relations, or subgraphs, for example smells, defects, patterns, or human ratings. The annotation scheme should align with intended use and should distinguish ``ground-truth'' labels from derived descriptors computed from the models (e.g., size category or automatically computed scores). Annotations can be obtained through expert judgment, crowd-based labeling, automated detectors, or hybrid approaches. For human labeling, clear guidelines and quality controls such as inter-rater agreement checks are important, because semantic judgments such as modeling quality or domain correctness are costly and subjective. To remain usable, annotations need explicit, machine-readable links to the underlying models and elements, for example via stable identifiers, trace links, or separate mapping files. The result of this stage is a base or processed dataset plus one or more annotation layers, which are often not uniform across datasets and therefore difficult to benchmark generically. For this reason, the initial benchmark in this work focuses rather on intrinsic model properties and treats annotation quality and coverage as a potential future extension.}

\ins{A third optional phase is \textbf{annotation and enrichment}, which introduces task-specific information layers on top of a base dataset. The annotation scheme should align with its intended use and distinguish ``ground-truth'' labels from derived descriptors computed by the models (e.g., model size, average element count). Annotations can be obtained through expert judgment, crowd-based labeling, automated detectors, or hybrid approaches. In general-purpose data annotation, crowd labeling has been shown to be useful when tasks are decomposed and quality controlled~\cite{SnowOJN08}, whereas conceptual model annotations often require domain and modeling expertise. Annotations can range from coarse model-level labels (e.g., domain categories) to fine-grained markings of elements, relations, or subgraphs (e.g., smells, defects, patterns, or human ratings). In principle, they may be even more fine-grained, e.g., attached to properties such as names, types, multiplicities, or layout attributes. However, such fine-grained annotation layers remain rare in current conceptual model datasets, which is one reason why generic annotation benchmarking is difficult today.}

\ins{Even when annotation benchmarking is left to future work, annotation layers should be designed so that they can be evaluated systematically. 
First, the annotation scheme should state the intended task, the annotation target and granularity, the meaning and admissible values of each label, inclusion and exclusion criteria, and representative positive and negative examples. 
Second, each annotation should have an explicit machine-readable link to its target, using stable model identifiers or, for finer granularities, stable element or relation identifiers, view identifiers, trace links, or separate mapping files. 
Third, annotation provenance should record the annotation method, annotator or tool, timestamp, data source, guideline version, and, where applicable, confidence. 
Finally, quality control should report coverage, missing or inconsistent labels, schema violations, duplicate or dangling annotation links, and, for human labeling, inter-rater agreement plus the procedure used to resolve disagreements. Established measures include Cohen's kappa for two raters~\cite{cohen1960coefficient} and Krippendorff's alpha for more general coding settings with multiple coders, missing values, or different measurement levels~\cite{krippendorff2018content}.}

\ins{ModelSet illustrates a common form of dataset enrichment through annotations~\cite{LopezIC22}. Its labels are attached to whole models, i.e., each Ecore or UML model receives a main \texttt{category} label that captures the application domain, and additional labels such as \texttt{tags}, \texttt{purpose}, \texttt{notation}, \texttt{tool}, \texttt{main-diagram}, and \texttt{confidence}. The labels are entered as structured \texttt{label: value} pairs and stored in a SQLite database through an Eclipse-based labeling tool. The tool implements GMFL, a greedy methodology for fast labeling where models are first ordered by the density of similar models, then the annotator inspects one unlabeled model and assigns a label, after which the tool retrieves ranked lists of similar unlabeled models to support a ``labeling streak'' in which several related models can receive the same category with less repeated cognitive effort. The process also includes quality control, where annotators review random samples of each other's labels and adjudicate disagreements.}

% For Ecore metamodels, similarity retrieval uses the MAR model search engine; for UML models, string attributes are indexed as text and retrieved with Whoosh. During inspection, the tool shows model trees, diagram-specific visualizations, outlines, source URLs, and related models from the same project, and it can asynchronously launch new searches from already labeled models to refine and extend the current streak. The annotation process was carried out by two experienced modelers, one for Ecore and one for UML. Quality control combined milestone discussions to align label usage with cross-validation: each annotator reviewed a random 25\% sample of the models labeled by the other annotator, disagreements were adjudicated by discussion, and the validation led to label changes for 3\% of Ecore models and 5\% of UML models. ModelSet is therefore a useful example of systematic model-level annotation, while also showing the cost of semantic labels that require modeling and domain expertise.

\subsubsection{Packaging and Use}
Once a dataset has been processed and optionally enriched, it is \textbf{packaged and used} in concrete tasks. Packaging typically includes documentation, dataset organization, and versioning, sometimes guided by FAIR principles~\cite{wilkinson2016fair}, and ideally accompanied by reproducible scripts and configurations for processing steps. Datasets are generally published in GitHub or Zenodo repositories. For actual tasks, researchers often transform the underlying models into different encodings, such as typed graphs, triples, token sequences, embeddings, or hybrid representations~\cite{ali2023encoding}. This operationalization is strongly task-dependent; for instance, a dataset used for model classification might require encoding the textual labels of entire models, whereas a dataset for next-element prediction might use local graph neighborhoods or triples. Task setup also requires careful protocol design, including splitting strategies (e.g., train/validation/test splits) that avoid leakage across duplicates and variants, and reporting choices that ensure comparable results (e.g., evaluation metrics). In this context, benchmarking dataset properties becomes essential, for example, construct coverage, size skew, vocabulary characteristics, label sparsity, duplication levels, and domain focus, because these factors materially affect outcomes and their interpretation.

Across all phases, model datasets typically evolve through feedback loops. Observed parse failures may motivate tighter acquisition filters or additional dependency collection. Discovered skew in model sizes or domains can trigger resampling or targeted acquisition to improve coverage. Weak lexical signals, for example, pervasive placeholder naming, may prompt filtering rules, additional enrichment, or revised task design. From this lifecycle perspective, benchmarking is not an additional terminal step but a cross-cutting evaluation layer. It can be applied after acquisition for triage, after processing for characterization and change tracking, prior to release for transparent reporting, and at task time for assessing dataset-task fit and threats such as low construct coverage. At the same time, not all lifecycle aspects can be meaningfully quantified uniformly, for example, licensing decisions or the semantic validity of expert annotations.

\section{Model Dataset Benchmarking Framework}\label{sec:framework}
In this section, we abstract from the generic characteristics of model datasets and from those of existing datasets reported in \autoref{sec:cm-datasets} to propose a model dataset benchmarking framework. First, we present a conceptual model for generic model dataset benchmarking (\autoref{sec:metamodel}). Afterward, we introduce an initial set of benchmarking quality dimensions and measures (\autoref{sec:quality-dimensions}).

\subsection{Conceptual Model for Model Datasets Benchmarking}
\label{sec:metamodel}

\begin{figure*}
    \centering
    \includegraphics[width=1\linewidth]{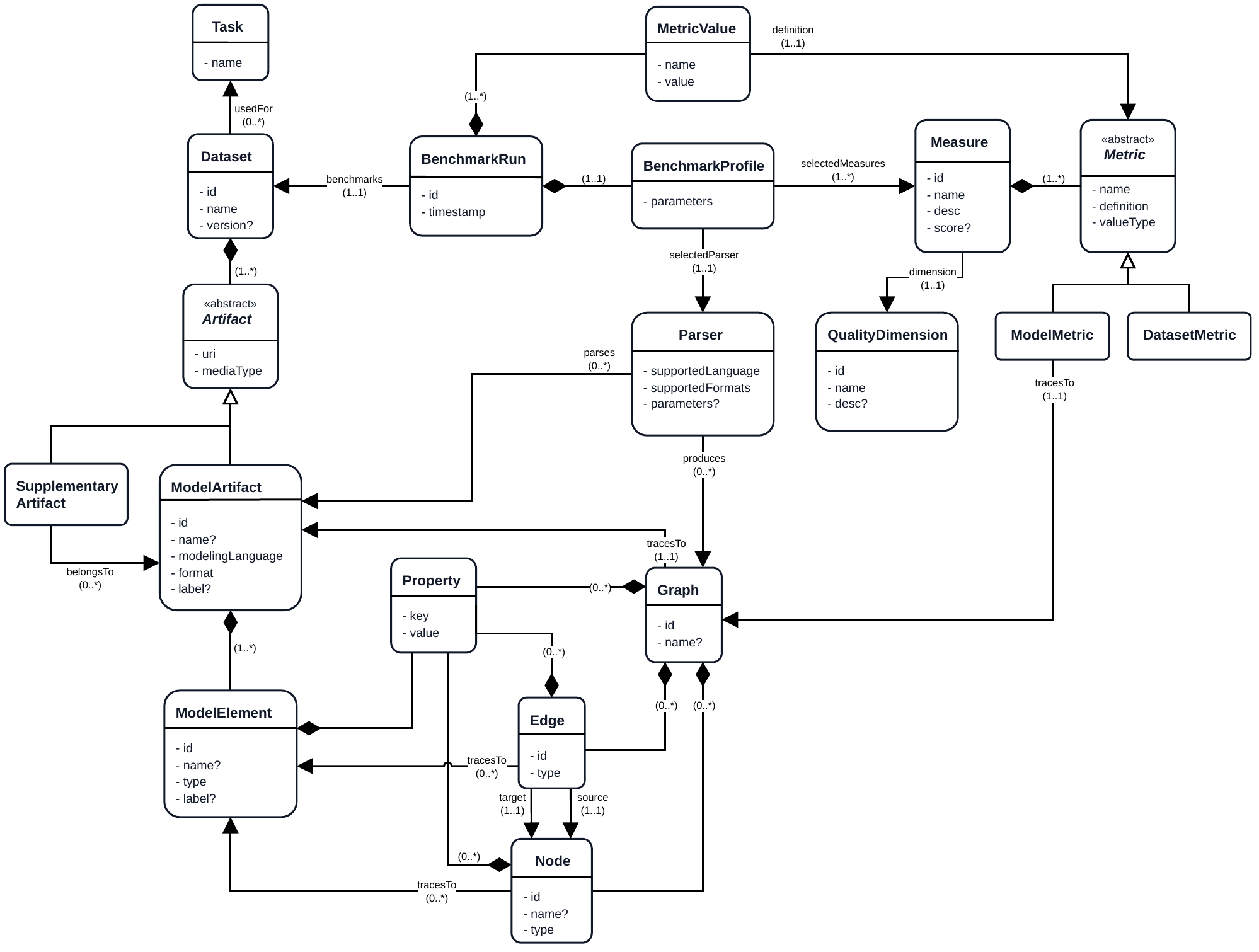}
    \caption{\del{Metamodel}\ins{Conceptual Model} for Model Dataset Benchmarking.}
    \label{fig:metamodel}
\end{figure*}

% Purpose and scope
\autoref{fig:metamodel} presents the \del{metamodel}\ins{conceptual model} that underpins our benchmarking approach. It defines the core entities and relationships required to benchmark model datasets in a reproducible, traceable, and extensible manner, while remaining independent of any particular modeling language and serialization format. \ins{The conceptual model guides the implementation of the prototype, but is not instantiated as a separate metamodel artifact. In the implementation, its concepts are realized through Python and JSON-based data structures.}

% Datasets and artifacts
In the metamodel, the \texttt{Dataset} is the central unit of benchmarking. A dataset is associated with identifying information (e.g., name and optional versioning to differentiate evolutions of the same dataset) and can be (conceptually) linked to one or more intended \texttt{Tasks}. In the current framework, tasks serve only as contextual information. The benchmarks themselves are formulated as task-agnostic descriptors of dataset characteristics, so that the resulting evidence can later be interpreted against different task requirements. This also keeps the framework open to future extensions that may introduce task-specific benchmark profiles or task suitability checks.

% Models
A dataset contains a collection of \texttt{Artifacts}, with a file URI and mediaType (e.g., \texttt{application/json}, or \texttt{image/png}). Artifacts are specialized into \texttt{ModelArtifact} and \texttt{SupplementaryArtifact}. Model artifacts represent models in some modeling language and serialization format \ins{with an optional label}. %, and they are the objects we are primarily interested in analyzing. 
Model datasets can include additional resources, which are captured via~\texttt{Supplemen\-taryArtifacts} to represent optional context (e.g., documentation, images, code) and potential dependencies (e.g., imported libraries, referenced fragments, modularized model parts). The metamodel represents \texttt{Model\-Elements} as instances of the language's abstract syntax, containing a unique ID, type (e.g., UML Class, Ecore EAttribute), and optionally a name \ins{and label}. Model elements can carry \texttt{Property} key–value pairs to represent additional language-specific properties (e.g., the source and target of a relationship, multiplicity, and stereotypes). This view on models and model elements provides the minimal structure required for traceability and for mapping elements into a shared intermediate representation.

% BenchmarkRun, Profile
\texttt{BenchmarkRun} and \texttt{BenchmarkProfile} then organize the benchmarking process itself. A benchmark run represents a single execution of the framework on a dataset, while a benchmark profile serves as a reproducibility unit that captures the configuration of a run, including parser selection, parameterization (e.g., lexical tokenizer settings), and enabled measures. The profile is crucial for comparability, as it makes explicit that benchmarking results depend on configuration choices, and it enables repeated runs on the same dataset under identical settings (i.e., reproducibility).

% IR, Parser
To compute metrics in a language-independent way, the metamodel introduces a graph-based Intermediate Representation (IR). A \texttt{Parser} transforms a model artifact into a concrete \texttt{Graph} representation while handling language- and format-specific details. The graphs consist of \texttt{Node} and \texttt{Edge} objects representing the model elements (traced back through their identifiers), each of which can carry additional property instances for language- and format-specific attributes. Thus, the parser is responsible for mapping model elements to nodes and edges. %, which introduces an important limitation: benchmark evidence is conditioned on the chosen parser implementation and its robustness. 
In this work, we treat parsing outcomes as benchmark evidence and transparently report failures, warnings, and which elements were loaded or skipped.

The overall model dataset benchmarking approach is organized around \texttt{QualityDimension}, \texttt{Measure}, and \texttt{Metric}. Quality dimensions provide conceptual groupings of benchmarking concerns (e.g., parsing, lexical quality, construct coverage, size). Measures refine a dimension into concrete aspects of interest (for example, Parse Status within Parsing or Label Presence within Lexical Quality). Optionally, a scalar score (in the range $[0,100]$) may be defined for a measure as an interpretation aid when this is meaningful (e.g., Parse Status can be given a score when giving different weights to successful/partial/failure parses), otherwise the measure is purely descriptive (e.g., Label Length for names cannot be scored as there is no notion of ``bad'' or ``good'' in regards to how many characters a label has). Measures are operationalized by one or more metrics. The metamodel distinguishes \texttt{ModelMetrics}, computed per model to support traceability and outlier inspection (e.g., how many nodes are contained in a model), from \texttt{DatasetMetrics}, computed as dataset-level aggregates from the model-level metrics (e.g., the number of nodes across all models in the dataset). \texttt{MetricValue} captures the computed result of a metric in the context of a specific benchmark run. Metric values may be simple numeric or categorical values, or more complex aggregated data types (e.g., maps of counts by type).

\begin{figure*}
    \centering
    \includegraphics[width=1\linewidth]{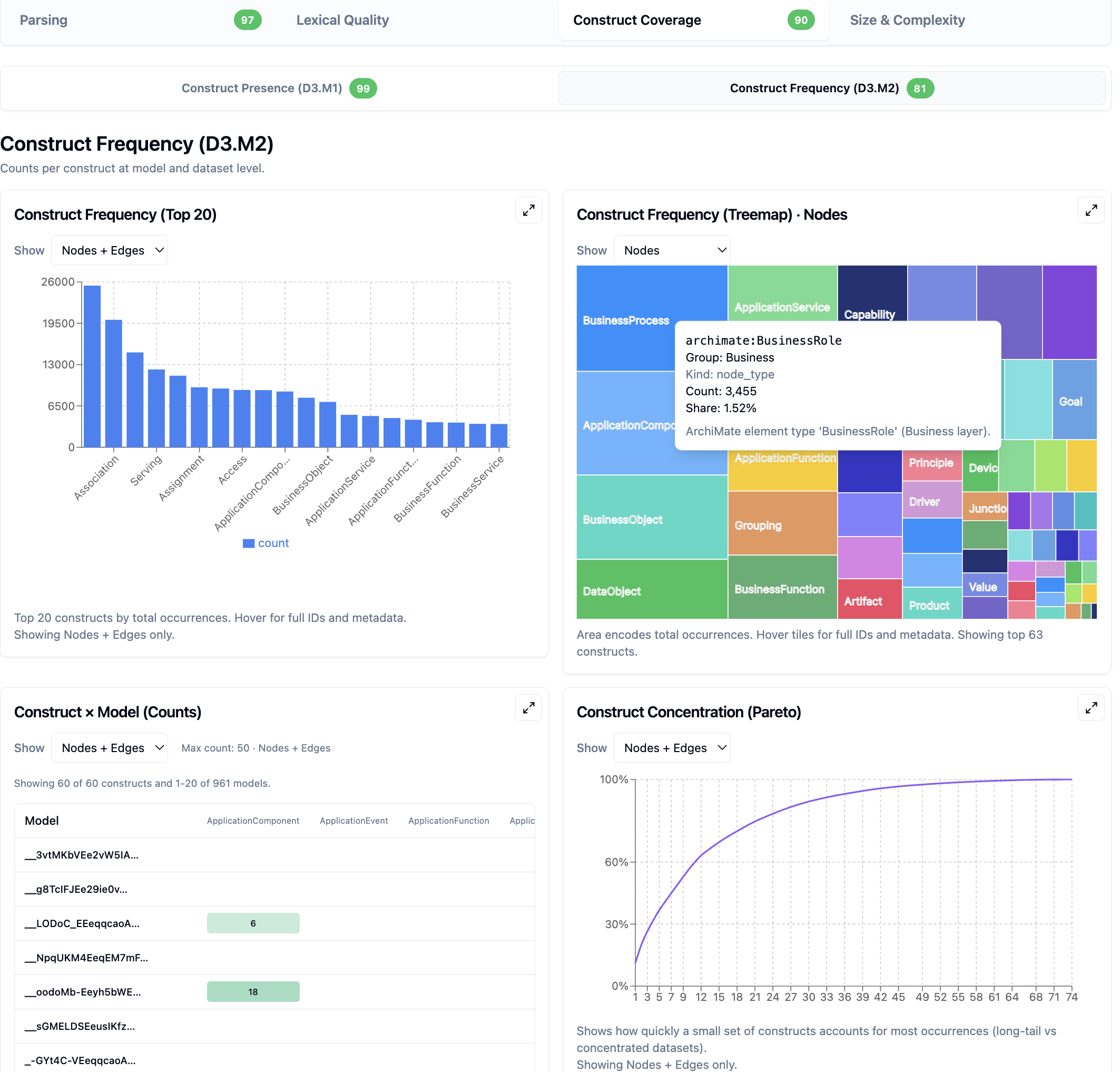}
    \caption{Web UI of the Benchmarking Prototype showing the Construct Frequency (D3.M2) View (excerpt).}
    \label{fig:tool-screenshot}
\end{figure*}

For interpretability, benchmark evidence must be presented in human-readable forms for analysis, such as histograms, tables, scatter plots, top-N lists, and coverage matrices. In our work, this is achieved through a projection step that derives report-ready aggregates from raw metric values, which are rendered in the tool's user interface (see \autoref{sec:prototype}). In the user interface, each measure has its own view where different visualizations are shown. \autoref{fig:tool-screenshot} shows an excerpt of the Construct Frequency (D3.M2) measure as rendered in the tool. We treat reporting as a separate presentation-oriented transformation layer on top of the metamodel, because report structures are inherently dependent on visualization choices, highly dynamic (e.g., filtering/sorting in tables, tooltips in plots), and can evolve independently of measurement semantics. This separation allows the framework to maintain stable metric definitions while supporting multiple (possibly evolving) reporting visualizations.

Overall, the \del{metamodel}\ins{conceptual model} is designed to make the benchmarking extensible and future-proof. New modeling languages and formats can be integrated by adding parsers and mappings to the Intermediate Representation. New benchmarking capabilities can be added by introducing additional quality dimensions, measures, and metrics. Finally, interpretation layers can evolve by refining scores or adding derived report projections, while the underlying metric evidence remains comparable across implementations and over time.
% Minimal changes might still happen to the metamodel over time, e.g., when supporting benchmarking multiple datasets (currently 1..1) or multiple languages within a dataset (also currently 1..1).

\subsection{Initial Quality Dimensions for Benchmarking}\label{sec:quality-dimensions}

\newcolumntype{P}[1]{>{\raggedright\arraybackslash}p{#1}}
% Ragged-right X column
\newcolumntype{Y}{>{\raggedright\arraybackslash}X}
% Centered fixed-width column
\newcolumntype{C}[1]{>{\centering\arraybackslash}p{#1}}

\begin{table*}[t]
\caption{Initial catalog of model dataset quality dimensions, measures, and intent.}
\label{tab:quality-dimensions}
\centering
\small
\setlength{\tabcolsep}{6pt}
\renewcommand{\arraystretch}{1.18}

\begin{tabularx}{\linewidth}{@{}P{2.2cm} P{4.5cm} Y C{0.9cm}@{}}
\toprule
\textbf{Dimension (D\textsubscript{x})} &
\textbf{Measure (D\textsubscript{x}.M\textsubscript{y})} &
\textbf{Intent} &
\textbf{Score} \\
\midrule

\multirow[t]{5}{2.6cm}{\textbf{D1 Parsing}} &
\parbox[t]{4.0cm}{\textbf{D1.M1 Parse Status}} &
\del{Quantifies the share of models that parse successfully, partially, or fail.} \ins{Quantifies the share of models for which the selected parser produces a complete, partial, or no usable intermediate representation.} &
\cmark \\
\cmidrule{2-4}

& \parbox[t]{4.0cm}{\textbf{D1.M2 Elements Loaded vs Skipped}} &
Assesses how much model content is dropped during parsing (skipped elements), overall and per model. &
\cmark \\
\cmidrule{2-4}

& \parbox[t]{4.0cm}{\textbf{D1.M3 Parsing Time}} &
Characterizes the distribution of per-model parsing times as a scalability indicator. &
\xmark \\
\cmidrule{2-4}

& \parbox[t]{4.0cm}{\textbf{D1.M4 File Size}} &
Describes model size on disk (source vs IR) and its variation across the dataset. &
\xmark \\
\cmidrule{2-4}

& \parbox[t]{4.0cm}{\textbf{D1.M5 Warnings}} &
Quantifies models that trigger parser warnings and the dominance of warning types. &
\cmark \\

\midrule

\multirow[t]{5}{2.6cm}{\textbf{D2 Lexical Quality}} &
\parbox[t]{4.0cm}{\textbf{D2.M1 Label Presence}} &
Measures label coverage where labels are expected and identifies where they are missing. &
\cmark \\
\cmidrule{2-4}

& \parbox[t]{4.0cm}{\textbf{D2.M2 Label Length}} &
Describes typical label lengths (characters/tokens) and the prevalence of very short or long labels. &
\xmark \\
\cmidrule{2-4}

%& \parbox[t]{4.0cm}{\textbf{D2.M3 Naming Convention Consistency}} &
%Assesses consistency of naming styles (case styles) within models and across the dataset. &
%\xmark \\

& \parbox[t]{4.0cm}{\textbf{D2.M3 Single vs Multi-Word Labels}} &
Quantifies the share of single-word vs multi-word labels and its variation across models. &
\xmark \\
\cmidrule{2-4}

& \parbox[t]{4.0cm}{\textbf{D2.M4 Lexical Diversity}} &
\del{Characterizes vocabulary diversity} \ins{Characterizes the vocabulary richness and repetition of extracted labels (e.g., total tokens, unique tokens, type-token ratio)} &
\xmark \\
\cmidrule{2-4}

& \parbox[t]{4.0cm}{\textbf{D2.M5 \ins{Label} Language \del{Usage}\ins{Distribution}}} &
\del{Characterizes language diversity} \ins{Characterizes the distribution of natural languages inferred from extracted labels (e.g., English, German, Spanish). This does not refer to the modeling language of the artifact}. &
\xmark \\

\midrule

\multirow[t]{2}{2.6cm}{\textbf{D3 Construct Coverage}} &
\parbox[t]{4.0cm}{\textbf{D3.M1 Construct Presence}} &
Determines which defined language constructs appear at all and how complete construct coverage is. &
\cmark \\
\cmidrule{2-4}

& \parbox[t]{4.0cm}{\textbf{D3.M2 Construct Frequency}} &
Analyzes how evenly constructs are used and whether construct usage is dominated by a few types. &
\cmark \\

\midrule

\multirow[t]{4}{2.6cm}{\textbf{D4 Size}} &
\parbox[t]{4.0cm}{\textbf{D4.M1 Model Size}} &
Captures structural model size (nodes, edges, elements) and its distribution. &
\xmark \\
\cmidrule{2-4}

& \parbox[t]{4.0cm}{\textbf{D4.M2 Degree}} &
Describes local connectivity patterns via average and median node degrees. &
\xmark \\
\cmidrule{2-4}

& \parbox[t]{4.0cm}{\textbf{D4.M3 Connectivity}} &
Assesses model fragmentation in terms of components and isolated nodes. &
\xmark \\
\cmidrule{2-4}

& \parbox[t]{4.0cm}{\textbf{D4.M4 Containment Depth}} &
Characterizes hierarchical depth and rootedness of containment structures. &
\xmark \\

\bottomrule
\end{tabularx}
\end{table*}

%A quality assessment framework for modeling  ecosystems~\cite{di2024amino}
%Quality assessment of ecore models~\cite{lopez2014assessing}
%\cite{mengerink2017automated}
%\cite{bertoa2010quality}
%\cite{babur2022samos}
%\cite{doan2022discovery}

This section instantiates part of the metamodel introduced in \autoref{sec:metamodel} with an initial catalog of quality dimensions and measures (which are currently already implemented in our benchmarking prototype (cf. Section~\ref{sec:prototype})). The catalog is intended as an incremental starting point rather than an exhaustive taxonomy. 

Existing work on software and model quality served as inspiration for our initial catalog of quality dimensions, measures, and metrics. Quality in software has been discussed extensively over the past 50 years (e.g., McCall Quality Model~\cite{cavano1978framework}, Boehm Quality Model~\cite{boehm1976quantitative}), culminating in the well-known ISO/IEC 9126 quality model~\cite{losavio2003quality}. Classic software quality models, such as ISO/IEC 9126 and its successors, decompose quality into characteristics and sub-characteristics, operationalized via measurable attributes (metrics). They emphasize that quantifiable attributes at design time can serve as early indicators of external qualities such as maintainability and reliability. This view is mirrored in object-oriented measurement, where structural metrics on classes (e.g., size, coupling, inheritance depth) are used to predict external quality attributes and to support early design decisions~\cite{genero2005survey,bajeh2020object}.

At the level of modeling languages, several works have systematized metrics and quality notions for, e.g., UML class diagrams, Ecore, and other model types, again with the goal of assessing internal properties (e.g., complexity, cohesion) as proxies for external qualities~\cite{genero2005survey,bertoa2010quality,budgen2011empirical,mathur2016empirical,lopez2014assessing,di2014mining,SmajevicEtal21KG4EASmells}. For models, quality frameworks typically organize metrics along quality types such as syntactic, semantic, and pragmatic quality, often building on software quality standards for terminology and characteristics~\cite{dalisay2018quality,krogstie1995defining,krogstie2006process}. Recent work has begun to extend these ideas from individual models to ecosystem and repository modeling. For example, AMINO defines a quality assessment framework for ecosystems of modeling artifacts, linking model- and metamodel-level metrics (e.g., maintainability, understandability, complexity, coverage) to higher-level quality attributes and providing visual analytics over entire repositories~\cite{di2024amino}. These approaches demonstrate that (i) internal, language-level metrics can be systematically organized into quality models, and (ii) quality evaluation can be lifted from single artifacts to collections of models. 

Our catalog in \autoref{tab:quality-dimensions} follows this tradition but adapts it to the specific goal of benchmarking model datasets. First, we restrict ourselves to mainly language-agnostic measures that can be computed from the graph-based IR introduced in \autoref{sec:metamodel}, with some language-specific extensions. Second, we define metrics that are meaningful at both the model and dataset levels, so they can support dataset comparison and evolution and characterize datasets (e.g., construct coverage and usage distributions). Finally, the current dimensions deliberately focus on basic prerequisites and descriptive characteristics that are broadly applicable across modeling languages. More specialized quality notions from the literature (e.g., maintainability, consistency, redundancy, or annotation quality) can be added as further dimensions and measures within the same metamodel as the benchmark framework matures.

\ins{The initial catalog is also informed by corpus-based analyses of modeling artifacts. For example, Tairas and Cabot~\cite{tairas2015corpus} analyze DSL corpora by counting metamodel-element instances and by measuring in how many models each element appears. They complement this with co-occurrence and clone analyses to support language-engineering decisions for Puppet and ATL. Babur et al.~\cite{babur2024language} perform a large-scale usage analysis of EMF metamodels on GitHub, reporting usage of metaclasses, features/relationships, attributes, annotations, and generics to derive feedback for the Ecore metamodeling language. We reuse this kind of corpus-level evidence, which is most visible in D3, where D3.M1 captures construct presence and coverage, while D3.M2 complements raw usage counts with an entropy-based balance score for comparing datasets.}

Our catalog organizes dataset benchmarking evidence into quality dimensions as conceptual groupings (e.g., parsing, lexical quality, construct coverage, size), and measures as concrete diagnostic questions within a dimension. Measures are operationalized through metrics that produce model-level and dataset-level evidence. For some measures, an additional score is provided as a compact orientation signal (normalized to $[0, 100]$), which should be interpreted as a heuristic summary of the underlying metrics rather than as task-independent ground truth. For metrics without a score, the measure remains purely descriptive and does not imply a notion of ``good'' or ``bad'' (e.g., file size, label length, model size). \autoref{tab:quality-dimensions} provides an overview of the current catalog, listing the quality dimensions and their measures, including the intent of each measure and whether a score can be computed.

For completeness, the full list of computed measures and metric schemas is maintained in the accompanying repository\footnote{\url{https://github.com/plglaser/cmbenchmark/blob/main/docs/MEASURE_CATALOG.md}}, as an exhaustive enumeration of the currently supported four dimensions, 18 measures, and 114 metrics would exceed the scope of this paper. However, to illustrate how measures are operationalized, \autoref{tab:d1m1-parse-status-metrics} details the metrics underlying the $D_1.M_1$ Parse Status measure. This measure distinguishes, at the model level, whether a model parses successfully (no warnings), partially (at least one warning), or not at all (failed). In case of a parsing failure, a diagnostic error message is recorded. At the dataset level, it aggregates counts of successful, partial, and failed parses, derives their shares, and defines a robustness score that penalizes partial and failed parses. In the prototype, these metrics are projected into a small set of report views; the reporting column in \autoref{tab:d1m1-parse-status-metrics} exemplifies typical projections, e.g., simple KPIs (total models, shares, score), a distribution chart over parse statuses, and per-model tables highlighting problematic artifacts and their error messages. Parse status also serves as a gatekeeper, since models that fail to parse are excluded from subsequent computations that require an IR. \ins{Parsing metrics should therefore be interpreted as processability and reproducibility evidence, not as a dataset quality assessment. For mature, curated datasets, high parseability is expected, and D1 may have limited discriminative power. However, for mined, newly discovered, or heterogeneous datasets, parsing diagnostics provide an essential first check before other measures can be interpreted reliably.} This also motivates treating parsing as a dedicated ``quality dimension'' and makes the parse status evidence critical for interpreting benchmarking results.

\newcommand{\metriccell}[1]{\begin{tabular}[t]{@{}l@{}}#1\end{tabular}}

\begin{table*}[ht]
\caption{D1.M1 Parse Status: metrics, datatypes, reporting, intended use, and interpretation.}
\label{tab:d1m1-parse-status-metrics}
\centering
\scriptsize
\setlength{\tabcolsep}{3pt}
\renewcommand{\arraystretch}{1.15}

\begin{tabularx}{\linewidth}{@{}P{2.65cm} C{0.85cm} C{1.05cm} P{2.15cm} Y Y@{}}
\toprule
\textbf{Metric} &
\textbf{Level} &
\textbf{Type} &
\textbf{Reporting} &
\textbf{Used for} &
\textbf{Interpretation} \\
\midrule

\metriccell{\texttt{n\_models}} &
Dataset &
Integer &
Aggregate KPI &
Denominator for all parsing status shares and indices. &
Context only. Larger means a larger evaluation base, not better quality by itself. \\

\midrule

\metriccell{\texttt{n\_success,}\\ \texttt{n\_partial,}\\ \texttt{n\_failed}} &
Dataset &
Integer &
Status distribution bar chart &
Absolute status distribution. &
More successes are better. More partials or failures indicate reduced processability. \\

\midrule

\metriccell{\texttt{share\_success,}\\ \texttt{share\_partial,}\\ \texttt{share\_failed}} &
Dataset &
Float &
Status distribution chart &
Normalized status distribution for comparability across datasets. &
Values in \([0,1]\). Higher success share is better, while higher partial or failed share is worse. \\

\midrule

\metriccell{\texttt{score =}\\ \texttt{((n\_success +}\\ \texttt{0.5*n\_partial)}\\ \texttt{/n\_models)*100}} &
Dataset &
Float &
Aggregate KPI, score badge &
Single robustness signal that discounts partial and failed parses. Summarizes D1.M1. &
Values in \([0,100]\). Higher indicates better parser robustness for the dataset. \\

\midrule

\metriccell{\texttt{parse\_status}\\ \texttt{$\in$ \{success,}\\ \texttt{partial, failure\}}} &
Model &
\metriccell{Enum/\\String} &
Per-model table &
Identifies problematic models and eligibility for downstream measures. &
\texttt{success} is best for downstream evidence. \texttt{partial} means usable but diagnostic. \texttt{failure} excludes the model from IR-based measures. \\

\midrule

\metriccell{\texttt{parse\_error\_msg}} &
Model &
\metriccell{String\\(optional)} &
Per-model table &
Diagnostics attached when parsing fails. &
Diagnostic only. Presence explains a failure and is not numeric quality evidence. \\

\bottomrule
\end{tabularx}
\end{table*}

\del{The current catalog is intentionally incomplete and is expected to evolve. Several relevant dimensions are not yet covered, e.g., redundancy/duplication/similarity, assessment of annotation quality and coverage, or additional established model quality notions from the literature (e.g., consistency, maintainability, or complexity proxies) with solutions we aim to integrate presented in e.g., [4,20,29,39,45]. Such extensions can be added as new dimensions, measures, and metrics while preserving the metamodel's concepts in Section 4.1. In this sense, the catalog should be viewed as the first version of a benchmark specification that can evolve as new datasets, tasks, and quality concerns emerge.}

\ins{The current catalog is intentionally incomplete and is expected to evolve. Redundancy/duplication/similarity are particularly important for dataset diversity and for preventing inflated results in downstream tasks. In the current version, however, we deliberately leave redundancy/duplication/similarity as a future dimension because meaningful near-duplicate detection is not trivial and depends on task-specific similarity notions (e.g., structural graph similarity, lexical similarity, hybrid approaches). Existing work on model clone detection and model cleansing pipelines~\cite{MartinezGC22,HeLH25,khalilipour2022categorization,Djelic2025pipeline,RattanBS13,babur2022samos,Lopez2022-search,storrle2010towards} provides a basis for integrating such measures in the future framework. Other relevant extensions include assessment of annotation quality and coverage or additional established model quality notions from the literature (e.g., consistency, maintainability, or complexity proxies). Such extensions can be added as new dimensions, measures, and metrics while preserving the metamodel's concepts, as described in \autoref{sec:metamodel}. In this sense, the catalog should be viewed as the first version of a benchmark specification that can evolve as new datasets, tasks, and quality concerns emerge.}

\section{Model Dataset Benchmarking Platform}\label{sec:prototype}
In the following, we describe our efforts to realize our model dataset benchmarking framework through an implemented benchmarking platform that can be swiftly integrated into scientific and data science workflows. We first present the requirements for such a platform in \autoref{sec:prototype:requirements}. Then, \autoref{sec:prototype:architecture} introduces an architectural view of our platform, followed by the pipeline stages in \autoref{sec:prototype:pipeline}.

\subsection{Platform Requirements}\label{sec:prototype:requirements}

%The aim of the following requirements elicitation is to motivate the main architectural decisions taken.
%\lola{I am commenting out the sentence above as it shows that the plaform was develop first and the the requirements created to "motivate", when it should have been done the other way around or alternatively following an agile methogology. Basically, I think it is best to not say anything to avoid criticism}
In the following, we describe the stakeholders and potential use cases we foresee for the platform, and we list its functional and non-functional requirements.

\ins{The requirements below are not intended as a universal requirement specification for all possible model dataset benchmarking applications. Rather, they define the scope of the first instance of the framework introduced in Section~\ref{sec:framework}. We derived the requirements from four sources: (i) the stakeholder goals of dataset curators and MDE researchers, (ii) the dataset characteristics and lifecycle issues discussed in Section~\ref{sec:cm-datasets}, (iii) recurring concerns in benchmarking and data-driven approaches, in particular reproducibility, transparency, comparability, extensibility, and explicit reporting of configuration and evidence, and (iv) lessons learned from iteratively implementing the framework and applying it to heterogeneous model datasets with different languages, serializations, parser behavior, and dataset organization. Since we are not aware of an established platform that benchmarks model datasets, we present these requirements as an initial specification informed by related work, stakeholder needs, and our own experience. The specification is expected to evolve as further datasets, modeling languages, and downstream tasks are considered.}

\subsubsection{Stakeholders \& Goals}
%\lola{I am changing Use cases -> Goals because in the context of requirement engineering, these are goals}
Stakeholders of the prototype are mainly \textit{(i) dataset curators}, who need to delineate and version datasets, detect problematic artifacts, and publish evidence about dataset characteristics, and \textit{(ii) MDE researchers}, who apply or develop computational MDE approaches (e.g., automated analysis, ML, LLMs, intelligent modeling assistance) on datasets and who need to select datasets that fit a task and to justify this selection with transparent, comparable signals (e.g., parseability, label regimes, construct coverage, structural properties, etc.). Across these groups, the central goals are to characterize a dataset, enable comparison of datasets under a consistent protocol, identify problematic models (e.g., outliers, models with parsing issues, structural anomalies), and export evidence that can be cited and reproduced in scientific publications.

\subsubsection{Functional Requirements}

\begin{description}[style=unboxed]
   \item[\textbf{FR1 Profile-driven benchmarking runs.}] Benchmark executions shall be configured via a single benchmark profile specifying dataset location, parser selection, enabled dimensions/measures, and relevant configuration parameters. This makes runs reproducible and enables dataset comparison.

   \item[\textbf{FR2 Dataset scanning and boundary control.}] The platform shall discover candidate model artifacts in a dataset directory and apply explicit inclusion/exclusion rules (through file name patterns, e.g., \texttt{*.ecore}) and basic filters (e.g., unreadable files, size constraints). This is necessary because publicly sourced datasets often include other model and non-model exports that need to be filtered out.

   \item[\textbf{FR3 Multi-language parsing via pluggable parsers.}] The platform shall support multiple modeling languages and model serialization formats through a uniform parser interface and registry, enabling incremental extension without redesigning the pipeline.

   \item[\textbf{FR4 Intermediate representation generation.}] Parsed models shall be mapped into a shared, typed-graph intermediate representation (IR) that acts as a common substrate for language-agnostic measure computation.

   \item[\textbf{FR5 Parsing diagnostics as first-class evidence.}] The platform shall record parse outcomes (success/partial/failure) and diagnostics (warnings, skipped elements, parsing time, file sizes) to make technical usability explicit for later dimensions.

   \item[\textbf{FR6 Measures at model and dataset level.}\\] For each enabled measure, the platform shall compute (a) model-level metrics for traceability and outlier inspection and (b) dataset-level metrics over the entire dataset for comparison. 

   \item[\del{\textbf{FR7 Configurable measure execution.}}]
   \del{Measures shall be selectively enabled and parameterized through the benchmark profile (e.g., lexical configuration and tokenizer settings), such that different research scenarios can be supported without requiring code changes.}

   \item[\textbf{\del{FR8}\ins{FR7} Report projection.}\\]
   The platform shall derive report-ready representations from raw measures (e.g., distributions, binned summaries, top-N tables) without changing metric semantics, enabling interpretation and downstream consumption by different frontends.

   \item[\textbf{\del{FR9}\ins{FR8} Dual orchestration modes with shared semantics.}\\]
   \del{
   The platform shall support both a CLI mode for batch/automation and a Web/API mode for interactive exploration, while invoking the same core services to guarantee identical results for the same profile and inputs.}
   \ins{The platform shall support both a CLI mode for scripting, CI integration, and reproducible batch experiments, and a Web/API mode for interactive exploration of diagnostics, outliers, and report visualizations. Both modes shall invoke the same core services to guarantee identical results for the same profile and inputs.}
   
   \item[\textbf{\del{FR10}\ins{FR9} Persisted artifacts for inspection and reuse.}\\] Each pipeline stage shall persist its artifacts (scan results, produced IRs, measures, reports) in a stable file layout to enable inspection, sharing, and recomputation of later stages without re-running the entire pipeline.
\end{description}

\subsubsection{Non-functional Requirements}

\begin{description}[style=unboxed]
   \item[\textbf{NFR1 Reproducibility.}\\] Given the same dataset and benchmark profile, repeated runs should yield the same results. All configuration and outputs must be explicit (i.e., persisted rather than ephemeral in-memory).

   \item[\textbf{NFR2 Transparency and explainability.}] The platform should expose failures, partial parses, warnings, and outliers rather than hiding them behind aggregated scores. Traceability from dataset-level summaries back to model-level evidence is required to support debugging and meaningful interpretation. \ins{This traceability is identifier-based, as model identifiers are preserved throughout scan results, parsing diagnostics, IRs, per-model metrics, and report views.}

   \item[\textbf{NFR3 Extensibility.}\\] Adding a new language/format should only require implementing a new parser and its IR mapping. Similarly, adding a new measure should be localized to the measurement layer, with only the reporting projection requiring extension when new derived views are desired.

   \item[\textbf{NFR4 Language-agnostic core with language-specific extensions.}\\] Measure computation should operate over the shared IR wherever possible without implicit language assumptions. Where language-specific semantics are unavoidable (e.g., construct catalogs or containment interpretation), they should be introduced explicitly and locally to prevent uncontrolled divergence across languages.

   \item[\textbf{NFR5 Scalability.}\\] 
   \ins{The platform should handle datasets of the order of thousands of models on a standard workstation. To make this requirement quantifiable and verifiable, we use 5,000 models as the acceptance-test threshold, reflecting the scale of many publicly available datasets.}
   \del{The platform should handle datasets of the order of thousands of models on a standard workstation, being the minimum 5,000 models.} % In our evaluation (see \autoref{sec:demonstration}), we demonstrate that the current implementation processes datasets of up to 5475 models within practical time and memory bounds.

   \item[\textbf{NFR6 Robustness under noisy inputs.}\\] The platform should degrade gracefully. Corrupted or unsupported artifacts must be recorded and isolated as evidence rather than causing the entire platform to crash.

   \item[\textbf{\del{NFR7 Portability and low operational overhead.}}] 
   \del{The platform should run locally without external services and without operational prerequisites such as database deployment. A file-based persistence model supports portability and simple packaging in replication artifacts.}

   \item[\textbf{\del{NFR8}\ins{NFR7} Separation of measurement and presentation.}\\] Metric computation must be independent of UI concerns. Reporting should be implemented as a derived projection layer, so that alternative frontends and exports remain possible.

   \item[\textbf{\del{NFR9}\ins{NFR8} Consistent semantics across interfaces.}] CLI and Web/API execution must invoke the same stage logic and produce equivalent outputs to avoid interface-dependent results.
\end{description}

\subsection{Platform Architecture}
\label{sec:prototype:architecture}

\begin{figure*}
    \centering
    \includegraphics[width=1\linewidth]{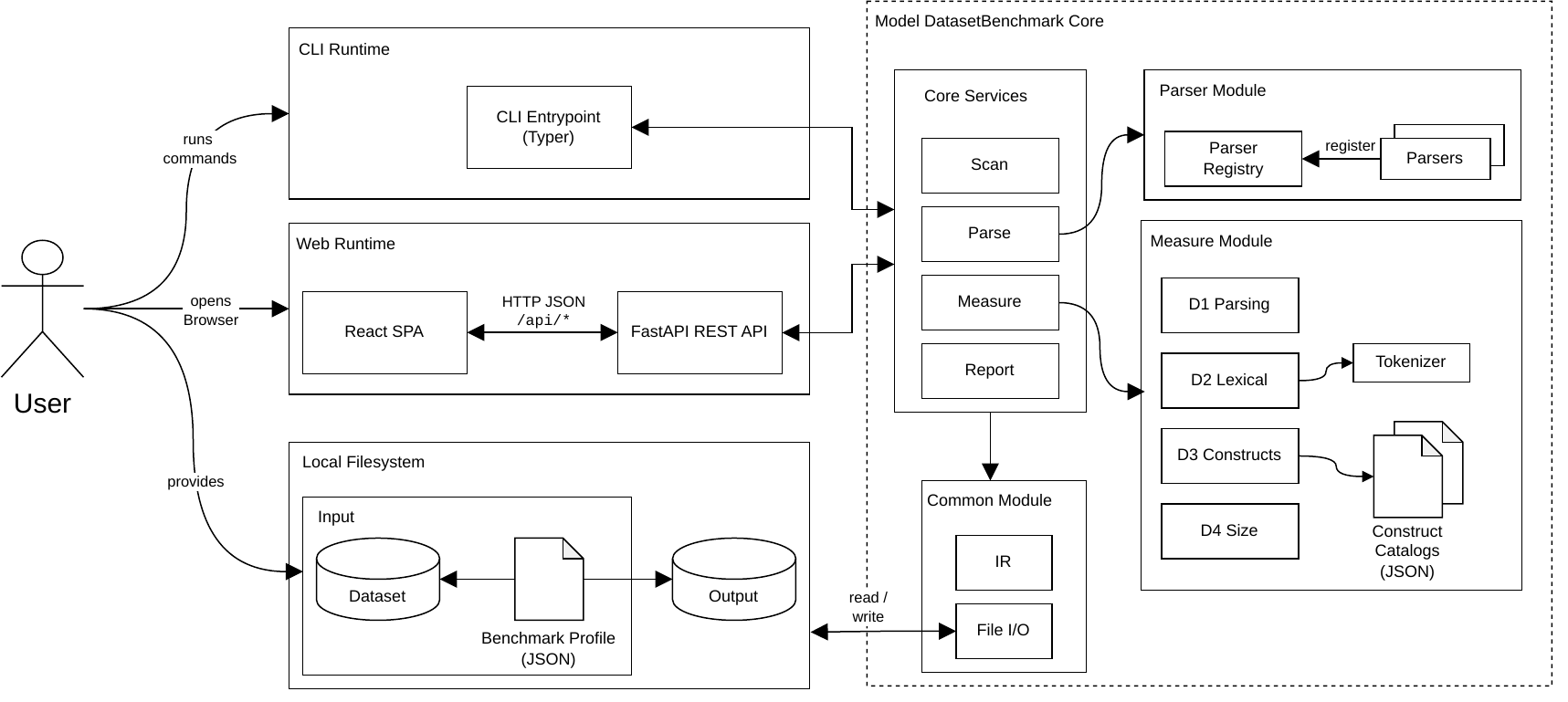}
    \caption{High-level Architecture of the Model Dataset Benchmarking Platform}
    \label{fig:architecture}
\end{figure*}

\autoref{fig:architecture} shows the architecture of the Model Dataset Benchmark platform. It consists of three main parts: \textit{(i)} user-facing runtimes, namely a \textit{CLI Runtime} and a \textit{Web Runtime}, \textit{(ii)} the \textit{Model Dataset Benchmark Core}, which implements all benchmarking logic and is shared by both runtimes, and \textit{(iii)} the local filesystem, as persistent store for datasets, benchmark profiles, and all produced artifacts (e.g., IR files, JSOM measures, JSON report). Both runtimes invoke the same core services, and all states are persisted as files rather than in a database, so that artifacts from each stage can be inspected, shared, and reused independently (\del{FR10}\ins{FR9}\del{, NFR7}). This realizes the main architectural choice of a file-based, profile-driven pipeline with shared core services, so that each benchmarking run is configured via a single benchmark profile (FR1) and remains reproducible given the same inputs (NFR1).

The upper-left of \autoref{fig:architecture} illustrates the runtime topology and user entry points. The \textbf{CLI runtime} is realized with Typer\footnote{\url{https://typer.tiangolo.com/}}, a library for building CLI applications, and provides a batch and scripting interface. It reads a benchmark profile and triggers the respective core services, either as individual stages or as an end-to-end run. This mode is intended for automation and reproducible experimentation, for example, when benchmarking multiple datasets under identical profiles or integrating benchmarking into scripted workflows (e.g., CI pipelines). At present, the CLI does not render visual reports, but it produces the same UI-ready JSON payload that the web runtime consumes, which could be extended in the future with command-line report generation (e.g., static plots).

The \textbf{Web runtime} is designed for interactive exploration and inspection of the benchmarking run results. The user interacts with a React Single-Page Application (SPA) in the browser, which communicates with a FastAPI\footnote{\url{https://fastapi.tiangolo.com/}} REST API using JSON. The frontend is built as a static bundle with Vite and served together with the API using uvicorn. The SPA offers a staged workflow aligned with the pipeline (Scan $\rightarrow$ Parse $\rightarrow$ Measure $\rightarrow$ Report) and renders visualizations from the derived artifacts of each stage, including the final report. The REST API itself remains thin, forwarding requests to the same core services used by the CLI, ensuring that stage semantics and output artifacts remain consistent across both orchestration modes (\del{FR9}\ins{FR8}, \del{NFR9}\ins{NFR8}).

The bottom part of \autoref{fig:architecture} shows the configuration and persistence model, which is based on the user's local filesystem. The user provides a benchmark profile in JSON format as the single configuration artifact \ins{(see \autoref{lst:benchmark-profile})}. The profile specifies the dataset location, the selected parser, which quality dimensions and measures are enabled (\del{FR7}\ins{FR1}) and their parameters (e.g., tokenizer configuration for lexical measures), and the output directory for generated artifacts. The dataset itself resides in a directory on the local filesystem, referenced from the profile. All results from the individual stages are written to the output directory specified in the profile, including scan results, IRs, computed measures, and the derived report payload (as described in \autoref{sec:prototype:pipeline}). This design simplifies deployment \del{(NFR7)}, avoids operational overhead from database infrastructure, and ensures that artifacts are easily inspectable, shareable, and suitable for archival, replication, and reproducible reruns (NFR1, \del{FR10}\ins{FR9}).

The right-hand side of \autoref{fig:architecture} depicts the Model Dataset Benchmark Core, which, on a high level, is organized into \textit{Core Services}, a \textit{Common Module}, a \textit{Parser Module}, and a \textit{Measure Module}. Each service follows the same interaction pattern: it reads its inputs from the filesystem, performs a stage-specific transformation, and persists its outputs back to the filesystem. Shared infrastructure is encapsulated in a common module to reduce duplication and which provides, e.g., file I/O utilities or typing information, including the IR. The IR implements the graph-based representation of the metamodel in \autoref{fig:metamodel} and serves as the uniform internal format for subsequent measurements (FR4, NFR4).

Language- and format-specific processing is isolated in the parser module, which constitutes the primary extensibility boundary for supporting additional modeling languages and serialization formats. A parser registry exposes available parser implementations, and each parser conforms to a uniform interface and registers itself for selection, enabling additional languages to be integrated without changing the surrounding pipeline (FR3, NFR3, NFR4). Parsers transform model artifacts into IR graphs and additionally emit parsing diagnostics, such as warnings, skipped elements, and timing information. Since benchmarking evidence depends on successful parsing and parser behavior, these diagnostics are treated as first-class outputs and directly support the parsing dimension (D1) and the requirement to expose parsing behavior as evidence (FR5, NFR2, NFR6).

\begin{figure*}[t]
    \centering
    \includegraphics[width=1\linewidth]{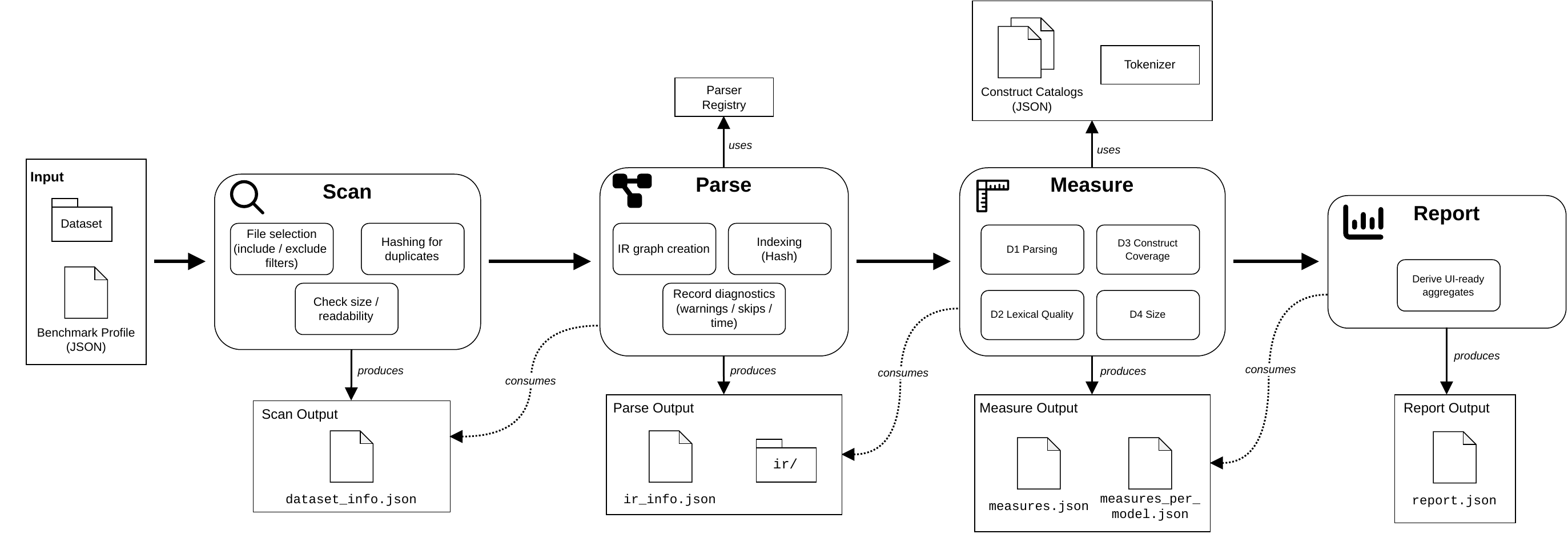}
    \caption{Model Dataset Benchmarking Pipeline Stages (bold horizontal arrows indicate the sequential execution order of the pipeline stages).}
    \label{fig:pipeline}
\end{figure*}

Finally, the measure module computes the benchmark evidence and implements the quality dimensions introduced in \autoref{sec:quality-dimensions}. Submodules correspond to the currently implemented dimensions (D1 Parsing, D2 Lexical Quality, D3 Construct Coverage, and D4 Size). They consume the IR and, where required, parsing diagnostics to compute model-level and dataset-level metrics (FR6). Measurement execution is configurable through the benchmark profile: measure groups can be enabled or disabled, lexical measures depend on a configurable tokenizer and label-selection settings, and construct measures depend on a language-specific construct catalog provided in JSON format. The module structure makes measures pluggable and independently evolvable: new dimensions or measures can be added as submodules without changing the surrounding architecture, while existing services and runtimes remain unchanged.

\subsection{Pipeline Stages}\label{sec:prototype:pipeline}
In the following, we briefly describe how the platform can be used in a concrete benchmarking scenario for a model dataset. As mentioned above, the description is structured along the phases of the benchmarking pipeline, i.e., Scan $\rightarrow$ Parse $\rightarrow$ Measure $\rightarrow$ Report.

\begin{figure}[h]
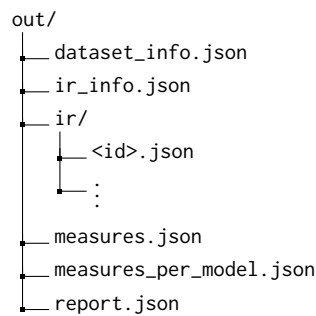

\dirtree{%
.1 out/.
.2 dataset\_info.json.
.2 ir\_info.json.
.2 ir/.
.3 \textless id\textgreater.json.
.3 \vdots.
.2 measures.json.
.2 measures\_per\_model.json.
.2 report.json.
}
\caption{Output Structure}
\label{fig:output-dir}
\end{figure}

\autoref{fig:pipeline} depicts the end-to-end model dataset benchmarking pipeline as a sequence of profile-driven stages with persisted artifacts. The pipeline takes a benchmark profile and a model dataset directory (referenced in the profile) as inputs. Each stage reads the outputs of the preceding stage and writes its own JSON artifacts to the profile's output path, so that later stages depend only on artifacts produced earlier and can be rerun independently (e.g., recomputing measures without reparsing). The resulting output structure is shown in \autoref{fig:output-dir}, consisting of \texttt{dataset\_info.json}, which captures scan results, \texttt{ir\_info.json} containing the parse results, \texttt{ir/} representing the IRs as one JSON file per model, \texttt{measures.json} and \texttt{measures\_per\_model.json} storing the raw metric values, and \texttt{report.json} holding the derived report payload.

% Benchmark Profile
The benchmark profile serves as the single configuration artifact and defines the reproducibility boundary of a run (FR1, NFR1). At a high level, it specifies (i)~the dataset location and file selection rules used during scanning (e.g., include/exclude patterns and size limits), (ii)~the parser language used in the parsing stage, (iii)~which measure groups are enabled (\del{FR7}\ins{FR1}) and any required parameters (e.g., tokenizer), and (iv)~the output directory in which all artifacts are persisted. \autoref{lst:benchmark-profile} shows a condensed example of a basic benchmark profile. The full schema and additional configuration options are provided in the accompanying repository~\cite{BenchmarkingPlatformRepo}.

\begin{lstlisting}[float, language=json,firstnumber=1, caption={Benchmark Profile (Excerpt)}, label={lst:benchmark-profile}]
{
  "name": "EAModelSet_Benchmark_01",
  "version": "1.0",
  "output_path": "./out",
  "scan": {
    "dataset_path": "./data/archimate-models",
    "include": ["*.archimate"],
    "exclude": ["**/tmp/**"],
    "size_limit_mb": 10
  },
  "parse": {
    "parser_language": "ArchiMate-Archi"
  },
  "measure": {
    "parse": {
      "enabled": true,
      "enable_d1_m1": true
      // other toggles omitted (enabled by default)
    },
    "lexical": {
      "enabled": true,
      "include_nodes": true,
      "include_edges": false,
      "label_attributes": ["name"],
      "tokenizer": { ... }
    },
    "constructs": { "enabled": true },
    "size": { "enabled": true }
  },
  "report": {}
}
\end{lstlisting}

\subsubsection{Stage 1: Scan} The scan stage identifies candidate model files in the model dataset directory and captures basic scan-level diagnostics. Its inputs are the dataset path and scan configuration from the benchmark profile. The stage performs recursive file discovery, applies include/exclude filters, and checks basic file accessibility constraints such as readability and optional size limits, thereby enforcing explicit dataset boundaries and filtering noisy artifacts (FR2, NFR6). 
For each remaining candidate, the scanner computes a SHA-256 hash of the file contents to detect exact duplicates. Only one representative per duplicate hash group is retained as a candidate for parsing. 

The scan stage produces the \texttt{dataset\_info.json} file, which records the resolved dataset root, scan parameters, summary totals (e.g., total files seen, candidates, unreadable, too-large files, and duplicate groups), counts by file extension, and the final list of candidate relative paths.
%\autoref{lst:scan-output} shows an example using a simplified dataset. 
Scan results are not benchmarked directly but serve as the first step in the explicit artifact trail, documenting which files were considered and why others were excluded, which is important for interpreting later measures (NFR2).

\subsubsection{Stage 2: Parse}\label{sec:prototype:parse}
The parse stage maps each candidate file into the graph-based IR and records parsing diagnostics. Its inputs are the candidate list from the previous scan step and the parser selection from the profile, both resolved through the parser registry. For each candidate artifact, the stage computes a deterministic identifier derived from the file content (via SHA-256 hashing) and invokes the selected parser to produce an IR graph. Source model elements are mapped to IR nodes and edges, while additional information can be retained as key–value attributes at the graph, node, and edge levels. This includes model-level metadata, where available (e.g., model version, author, documentation), as well as language-specific properties (e.g., whether a class is abstract or the relationship multiplicities).

\ins{Parsing denotes the process of traversing a serialized artifact, interpreting its structure according to a defined syntax, and constructing an internal representation of that structure.
%loading or traversing the serialized representation, resolving the information needed by the parser, and projecting the result into the IR. 
We distinguish two implementation strategies. Metamodel-based parsers use frameworks such as PyEcore\footnote{\url{https://github.com/pyecore/pyecore}} to parse a model (e.g., a UML model using the UML metamodel) or a metamodel (e.g., an Ecore metamodel) into an object graph and then derive IR nodes and edges from that graph. This reduces the amount of syntax-specific handling (e.g. XML) and delegates aspects such as namespace handling, containment traversal, and typed feature access to the framework where possible. On the other hand, serialization-oriented parsers, by contrast, read XML, JSON, or text directly and construct the IR from explicit parser-specific extraction rules. This strategy is useful when no metamodel is available, when a possibly more lightweight, curated language-specific graph is preferable, or when avoiding a dependency on an external library (i.e., PyEcore, which is still actively developed) gives more control over the emitted IR. The trade-off is that concept and reference coverage depend on the mappings implemented by the parser, making these parsers easier to customize but also more error-prone. For UML, we provide both a PyEcore-based XML/XMI parser, which produces a more verbose, lower-level graph (since each metamodel concept is mapped to the IR), and a handcrafted XML/XMI parser, which maps a subset of supported UML concepts to a more curated semantic graph.}

\ins{The currently supported parsers are:
\begin{itemize}
    \item \texttt{Ecore}: loads \texttt{.ecore} metamodel files with PyEcore\footnote{Note that instance models conforming to an Ecore metamodel are not supported}.
    The parser identifies the root packages, traverses their contained EObjects, and projects them into the IR. Selected contained EObjects are represented as nodes, while containment, inheritance, reference, and type relations are represented as edges. References to external datatypes or classes are handled on a best-effort basis, i.e., when the referenced resource can be resolved or the target can be recovered from the serialized reference, the parser materializes the target, otherwise it records a warning and preserves the remaining parse result.
    Ecore annotations, including OCL constraints, are retained as raw metadata in the IR, for example in \texttt{node.data[``annotations'']}, \texttt{edge.data[``annotations'']}, or the top-level \texttt{ir.data[}
    \texttt{``modelAnnotations'']}. However, they are merely preserved and not validated or used in the subsequent pipeline stages.
    \item \texttt{UML-XML-PyEcore}: registers the Eclipse UML2 metamodel\footnote{\url{https://github.com/eclipse-uml2/uml2}}, including relevant namespace aliases, pathmap locations, and primitive libraries, before loading UML XML/XMI files with PyEcore. It traverses the loaded UML object graph, creates IR nodes for loaded contained UML objects, and creates IR edges for loaded non-derived, non-transient references whose targets are also available. This parser favors broad metamodel-driven loading at the cost of a more verbose lower-level IR.
    \item \texttt{UML-XML}: provides the handcrafted UML XML/XMI parser. It reads the XML tree directly, builds an \texttt{xmi:id} index, dispatches supported UML concepts to dedicated handlers, and materializes selected containment, ownership, and reference edges. This parser produces a more curated semantic graph and gives full control over how UML concepts are represented, but its coverage is limited to the concepts and references for which mappings and handlers have been implemented.
    \item \texttt{ArchiMate-Archi}: parses the XML format used by the Archi tool. It starts at the root, traverses element folders to create nodes, traverses the relations folder to create edges, normalizes ArchiMate element and relationship types (e.g., aligning deprecated types with newer ones), and records warnings for unresolved relationship endpoints.
    \item \texttt{BPMN-Signavio-JSON}: parses BPMN models exported in the Signavio JSON format. It checks for a top-level \texttt{BPMNDiagram}, recursively collects shapes, separates node-shaped stencils from edge-shaped stencils, resolves edge endpoints from incoming/outgoing references and target fields, and emits BPMN nodes, flow edges, and containment edges.
\end{itemize}}

\del{This stage classifies parsing outcomes into \texttt{success} (IR constructed with no warnings and no skips), \texttt{partial} (IR constructed with at least one warning and/or skips), or \texttt{failure} (IR could not be constructed) and records diagnostics such as warnings, skipped elements, parsing time, and file sizes (source vs IR). The stage writes one IR file per successfully parsed model under \texttt{\textless id\textgreater.json} and the file \texttt{ir\_info.json}, which aggregates dataset-level parse totals and provides a mapping from model identifiers to relative paths (index) together with per-model diagnostics. }

\ins{This stage classifies parsing outcomes into \texttt{success} (a non-empty IR is constructed with no warnings and no skips), \texttt{partial} (a non-empty IR is constructed with at least one warning and/or skip), or \texttt{failure} (no IR can be constructed). Thus, parse status measures processability under the selected parser and profile. It should not be interpreted as a certificate that the original model satisfies all metamodel conformance rules or language-specific well-formedness constraints. Parser implementations may perform, e.g., format checks, ID lookup, reference resolution, URI mapping, and compatibility adaptations where needed, and they record diagnostics when content is unresolved, adapted, unsupported, or skipped. The diagnostics also include parsing time and file sizes (source vs IR). The stage writes one IR file per successfully parsed model under \texttt{\textless id\textgreater.json} and the file \texttt{ir\_info.json}, which aggregates dataset-level parse totals, per-model diagnostics, and an index. This index is a JSON object mapping each generated deterministic IR identifier string to the corresponding source artifact path string relative to the dataset root (e.g., \texttt{\{``9f2a1c4b8d7e6a30'': ``models/order.ecore''\}}). The index is also used to trace benchmark results to source files.}

\del{Parse status acts as a gatekeeper for downstream stages, as models that fail to parse cannot contribute to measures that rely on IR-based analysis, and partial parses may reduce construct coverage or distort other statistics. The platform currently supports parsers for ArchiMate models stored in the Archi tool format and for Ecore models (using pyecore).}

\ins{Parse status also acts as a gatekeeper for downstream stages, as models that fail to parse cannot contribute to measures that rely on IR-based analysis, and partial parses may affect measures such as construct coverage, size, connectivity, or lexical statistics because these measures are computed over the IR elements that were actually materialized. Comparisons of measures across datasets are strongest when the same parser version, benchmark profile, and construct catalog are used. Otherwise, observed differences may partly reflect parser mapping choices rather than properties of the original datasets. Consequently, parser selection and parser diagnostics are made explicit as benchmark evidence. These diagnostics can also support manual repair of models, since warnings can be traced back to the affected model, and, where applicable, to affected elements through identifiers reported in the warning message. However, the richness and actionability of these warnings depend on the individual parser implementation, including whether a condition is reported at all and how much context the diagnostic message contains.}

\subsubsection{Stage 3: Measure} The measure stage computes the benchmark measures defined in \autoref{sec:quality-dimensions} and produces both dataset-level and model-level metrics. Its inputs are the IR files in \texttt{ir/}, the parse index and diagnostics in \texttt{ir\_info.json}, and the measurement configuration in the benchmark profile. Parsing measures (D1) are computed directly from \texttt{ir\_info.json}, aggregating parse statuses, skips, warnings, time, and file sizes. The remaining dimensions operate on the IR graphs and use the index for traceability back to source artifacts. Lexical measures (D2) analyze labels on selected IR entities according to the lexical configuration (e.g., node vs edge inclusion and label attribute selection) and apply a configurable tokenizer. For language detection (D2.M5), lingua-py\footnote{\url{https://github.com/pemistahl/lingua-py}} is used. Construct coverage measures (D3) compare observed IR types against a language-specific construct catalog to quantify presence and utilization (frequency) patterns. Construct catalogs used in D3 are language-specific JSON resources that map IR types to logical constructs. This indirection allows construct coverage measures to be extended without modifying the core pipeline, while isolating language-specific semantics in explicit resources (NFR4, NFR3). 

\autoref{lst:construct-catalog-example} shows an example construct catalog for a few Ecore constructs. The \texttt{kind} together with the \texttt{match\_type} is used to match IR nodes/edges, and the \texttt{match\_data\_equals} can further refine constructs by checking the \texttt{data} attributes of nodes/edges. The \texttt{meta} attribute allows the addition of contextual information to each construct. Finally, size measures (D4) compute structural graph characteristics, including node/edge counts, degree distributions, connectivity, and containment depth. 

The stage aggregates results into two persisted artifacts: \texttt{measures.json} stores dataset-level metrics, while \texttt{measures\_per\_model.json} stores corresponding per-model metrics, keyed by the IR identifier, thereby providing both model-level and dataset-level evidence for analysis and outlier inspection (FR6, NFR2). These files form the evidence layer of the model dataset benchmark as they preserve the computed raw metric values independently of any particular visualization.

\subsubsection{Stage 4: Report} The report stage transforms raw measures into a derived payload that the Web UI can render directly. It aggregates metrics into chart series, histograms, tables, and score badges without altering their semantics, providing a report projection layer on top of the evidence (\del{FR8}\ins{FR7}, \del{NFR8}\ins{NFR7}). Its inputs are \texttt{measures.json} and \texttt{measures\_per\_model.json} from the previous stage. For each measure, the report builder derives visualization-friendly structures such as binned histograms, scatter series, top-N tables, and matrices. Multiple report projections may exist for the same underlying metric values, reflecting that different visualizations can serve different analytical goals. 

The output of this stage of the pipeline is serialized into the \texttt{report.json} file, which provides a consolidated payload designed for direct consumption by the Web UI. The platform currently consumes this derived payload by React components that implement individual visualizations for each report object. Separating the evidence layer (measures) from the report projection ensures that new visualizations can be easily added or that CLI-based plotting can be introduced in the future.

\begin{lstlisting}[float, language=json,firstnumber=1, caption={Construct Catalog Example (excerpt and simplified)}, label={lst:construct-catalog-example}]
{
  "language": "Ecore",
  "constructs": [
    {
      "id": "ecore:EPackage",
      "kind": "node_type",
      "match_type": "EPackage",
      "match_data_equals": {},
      "meta": {
        "group": "metaclass"
      },
    },
    {
      "id": "ecore:EClass_Abstract",
      "kind": "node_type",
      "match_type": "EClass",
      "match_data_equals": {
        "abstract": true
      },
      "meta": {
        "group": "modifier"
      }
    },
    {
      "id": "ecore:EReference",
      "kind": "edge_type",
      "match_type": "Reference",
      "match_data_equals": {},
      "meta": {
        "group": "metaclass"
      }
    },
\end{lstlisting}

\section{Benchmarking existing Model Datasets}\label{sec:demonstration}
This section demonstrates the benchmarking platform on three publicly available model datasets with the goal of (i)~validating the end-to-end technical feasibility of the pipeline in processing heterogeneous model corpora, and (ii)~characterizing model dataset properties along the four implemented benchmark dimensions (D1–D4). This demonstrates the platform's ability to produce descriptive statistics for datasets rather than formal hypothesis testing. Accordingly, all reported values are conditional on the concrete parser implementations and their mapping into the IR. 

\subsection{Experimental Setup}\label{sec:experimental-setup}
The evaluation uses three of the publicly available model datasets reported in \autoref{sec:existing-datasets}. They differ in language, sourcing process, and level of curation. These are: EA ModelSet~\cite{Glaser25-EAModelset}, ModelSet~\cite{LopezIC22}, and AtlanMod Zoo\footnote{\url{https://github.com/AtlanMod/atlantic-zoo/}}.

EA ModelSet is a collection of ArchiMate models sourced from public repositories, including GitHub and GenMyModel, complemented by additional web sources and community contributions. For this study, all files in the Archi tool storage format (\texttt{*.archimate}) were extracted, yielding 961 candidate models after scan filtering. 
ModelSet~\cite{LopezIC22} is a large mined corpus of UML and Ecore software models. From this dataset, we selected all 5,475 Ecore metamodels (originally mined from GitHub) in the \texttt{*.ecore} file format.
AtlanMod Zoo is a curated repository of Ecore metamodels intended as shared experimental material. The maintainers explicitly note heterogeneous provenance (researchers and students) and therefore do not guarantee uniform quality or current relevance. The benchmark run included 304 candidate \texttt{*.ecore} files from the repository. Across all three datasets, no extra cleaning or manual normalization was applied beyond selecting the relevant file extensions. The intention is to evaluate the corpora as they are typically encountered during download.

These three datasets were chosen because together they span several relevant axes: mined versus curated artifacts, ArchiMate versus Ecore, and small/medium versus large corpora. EA ModelSet and ModelSet expose the prototype to noisy, heterogeneous material where modeling practices and tool versions vary, whereas AtlanMod Zoo provides a contrast in the form of a smaller, curated set of metamodels.

All runs are configured via benchmark profiles, and, for comparability, the configuration is kept as similar as possible across datasets. The scan stage uses simple include patterns (\textit{*.archimate} for EA ModelSet, \textit{*.ecore} for ModelSet and AtlanMod Zoo) without size limits or additional exclusion rules. Parser selection is dataset-specific: EA ModelSet is processed with the ArchiMate parser, whereas ModelSet and AtlanMod Zoo are processed with the Ecore parser, implemented in pyecore. All four dimensions D1–D4 and all associated measures are enabled in the profile. For the lexical dimension, we only used the \texttt{name} attribute label of nodes in the IR. A shared tokenizer is then applied to split on punctuation and camel case, trim whitespace, retain numbers, and ensure no lowercase letters. The concrete JSON benchmark profiles used for the experiments are provided in the accompanying repository\footnote{\url{https://github.com/plglaser/cmbenchmark/tree/main/profiles}}.

All experiments were executed on macOS 15.7.3 on an Apple M4 Pro machine with 24 GB RAM, using Python 3.13.5. The platform's web runtime was used, which executes the same core pipeline stages as the CLI and produces the persisted JSON artifacts. Each model dataset was run once through the full pipeline using its corresponding benchmark profile. The quantitative results reported in the following subsections are taken from the generated \texttt{measures.json} files of the measure stage.

\begin{table*}[ht]
\centering
\caption{Summary of D1 Parsing Metrics}
\small
\setlength{\tabcolsep}{3pt}
\renewcommand{\arraystretch}{1.15}

\begin{adjustbox}{max width=\textwidth}
\begin{tabularx}{\textwidth}{lP{1.0cm}P{1.6cm}P{1.5cm}P{1.3cm}P{1.5cm}P{2.8cm}P{0.9cm}llll}
\toprule
\textbf{Dataset} &
\textbf{Cand.} &
\textbf{Success} &
\textbf{Partial} &
\textbf{Failure} &
\textbf{Models w/ Skips} &
\textbf{Elements Loaded / Skipped} &
\textbf{Warn.} &
\textbf{M1} &
\textbf{M2} &
\textbf{M5} &
\textbf{D1} \\
\midrule
EA ModelSet &
961 &
900 (93.7) &
61 (6.3) &
0 (0) &
0 (0) &
229,855 (0) &
556 &
97 & 100 & 94 & 97 \\
ModelSet &
5475 &
5000 (91.3) &
400 (7.3) &
75 (1.4) &
59 (1.1) &
1,009,192 (422) &
803 &
95 & 100 & 93 & 96 \\
AtlanMod Zoo &
304 &
296 (97.4) &
4 (1.3) &
4 (1.3) &
3 (1.0) &
88,832 (5) &
6 &
98 & 100 & 99 & 99 \\
\bottomrule
\end{tabularx}
\end{adjustbox}
\label{tab:d1-results}
\end{table*}

The analysis is guided by four evaluation questions that structure interpretation across the measurement dimensions:

\begin{itemize}
    \item \textbf{Q1 - Technical usability (D1)}: Can the models in a dataset be parsed reliably and consistently with the prototype, and what kinds of issues (warnings, skipped elements, failures) occur?
    \item \textbf{Q2 – Lexical signals (D2)}: How usable are the element labels for text-based tasks in terms of presence, length, diversity, and language distribution?
    \item \textbf{Q3 – Construct representativeness (D3)}: Which language constructs are actually found in the datasets, and how balanced is their usage?
    \item \textbf{Q4 – Structural substance and variation (D4)}: What is the size distribution and graph structure of the models, in terms of scale, density, connectivity, and containment hierarchy?
    % \todo{Maybe leave Q5 out? Not discussed in too much detail}
    % \item \textbf{Q5 – Cross-dataset contrasts (D1–D4 combined)} In which respects do the datasets differ in ways that could bias or limit downstream evaluations?
\end{itemize}

The remainder of this section answers these questions, dimension by dimension, using the same metrics across all three model datasets. This enables a consistent comparison.

\subsection{D1. Parsing}
The parsing dimension (D1) captures the technical robustness of the platform's parsers across each model dataset. \autoref{tab:d1-results} summarizes the parsing-related benchmarking results (D1) across the three datasets. In our setup, a model parsing is classified as \textit{success} if an IR could be constructed without warnings or skips, as \textit{partial} if an IR was built but the parser raised warnings and potentially skipped elements, and as \textit{failure} if no IR could be created. Thus, ``partial'' reflects successful IR generation with diagnostics, not necessarily a formally partial or inconsistent model. Overall, the datasets are largely parseable with the selected parsers, yielding high scores and negligible information loss due to skipped elements. The remaining differences mainly concern the prevalence and nature of parser warnings, which serve as indicators of portability issues, tool- or dialect-specific idiosyncrasies, and potential deviations from expected modeling practices.
% More expected in large mined datasets (like ModelSet) as there is lots of variety

\textbf{Parse status and robustness.} The main indicator for technical usability is the distribution of parse outcomes (D1.M1). EA ModelSet contains 961 candidate models, of which 900 (93.7 \%) parse cleanly and 61 (6.3 \%) are classified as partial. There are no failures, yielding a D1.M1 score of 97. ModelSet is larger and slightly more heterogeneous, as of 5475 candidates, 5000 (91.3 \%) parse cleanly, 400 (7.3 \%) are partial, and 75 (1.4 \%) fail completely, resulting in a D1.M1 score of about 95. AtlanMod Zoo, which is manually curated, shows the highest robustness with 296 successful parses (97.4 \%), 4 partials (1.3 \%), and 4 failures (1.3 \%), corresponding to a D1.M1 score of 98. For subsequent dimensions, only models with a valid IR are considered (961 for the EA ModelSet, 5400 for the ModelSet, and 300 for the AtlanMod Zoo).

\textbf{Skips and warnings as quality signals.} D1.M2 and D1.M5 refine this picture and provide a more fine-grained view by distinguishing between benign warnings and warnings associated with actual content loss (skips). In all three datasets, the number of skipped elements is negligible. EA ModelSet has no skipped elements, ModelSet skips 422 out of more than one million loaded elements, and AtlanMod Zoo skips only 5 out of almost 90,000 elements. This yields D1.M2 scores effectively at 100 for all three datasets and indicates that, even for partial parses, almost all model content is preserved in the IR. 

\ins{The five skips in AtlanMod Zoo are duplicate reference edges occurring in three models. In each case, the first occurrence was retained in the IR and only the repeated edge was skipped, so no unique information was lost. The skips in ModelSet have a different character: 418 of the 422 skipped elements are generic Ecore references, which the parser currently cannot handle. The parser preserves the surrounding elements but does not currently materialize these parameterized reference targets as IR edges. Thus, there is a small loss of generic-reference information but not a substantial loss of model content. The remaining four skips are references whose endpoints could not be resolved due to unavailable external Ecore dependencies.} 

Warnings show clearer differences. In EA ModelSet, 61 models (6.3 \%) trigger warnings, and all are of type \texttt{UNRESOLVED\_REFERENCE} (556 total). Manual inspection suggests that a recurring pattern is the use of relationships as endpoints rather than as elements. While this can be resolved by additional scheduling logic in the parser in the future, this modeling style is legal in ArchiMate but unusual; thus, we intentionally retain it as a warning. Importantly, the relationships are retained in the IR with their original endpoints, which explains why warnings do not translate into skips for this dataset.

In AtlanMod Zoo, only 4 models (1.3 \%) produce warnings, most of them \texttt{DUPLICATE\_ID} (5 total) and one \texttt{UNRESOLVED\_REFERENCE}. Here, the warning corresponds to a concrete integrity issue at the serialization level, such as duplicated edge identifiers or duplicated references with identical endpoints. This warning type can directly lead to skipping an element in order to preserve a consistent IR, as reflected in the number of skipped elements.

\begin{table*}[t]
\centering
\caption{Summary of D2 Lexical Quality Metrics}
\label{tab:d2-summary}
\small
\setlength{\tabcolsep}{3pt}
\renewcommand{\arraystretch}{1.15}

\begin{adjustbox}{max width=\textwidth}
\begin{tabularx}{\textwidth}{lP{2.1cm}P{2.7cm}P{2.6cm}P{1.5cm}P{1.35cm}P{1.65cm}X}
\toprule
\textbf{Dataset} &
\textbf{Label presence (\%)} &
\textbf{Median length (chars / tokens)} &
\textbf{Median single-word (\%)} &
\textbf{Total tokens} &
\textbf{Vocab size} &
\textbf{English usage (\%)} &
\textbf{Distinct languages} \\
\midrule
EA ModelSet &
99.6 &
16 / 2 &
18 &
342,486 &
44,356 &
59.7 &
25 \\
ModelSet &
100 &
7 / 1 &
57 &
593,637 &
19,366 &
89.6 &
36 \\
AtlanMod Zoo &
100 &
8.5 / 1 &
55 &
50,757 &
6,435 &
97 &
6 \\
\bottomrule
\end{tabularx}
\end{adjustbox}
\end{table*}

ModelSet shows the most diverse warning profile, as expected for a large, mined dataset that aggregates heterogeneous toolchains and metamodel versions. Parsing the ModelSet yielded 293  \texttt{UNRESOLVED\_REFERENCE} warnings (often to missing external files), and 4 warnings indicating dangling references (\texttt{MISSING\_EDGE\_ENDPOINT}). Furthermore, \texttt{COMPATIBILITY\_ADAPTATION} warnings occurred 88 times, relating to unsupported Ecore constructs (e.g., the \texttt{eKeys} attribute throws an error in the underlying pyecore parser). Finally, 418 warnings of type \texttt{UNSUPPORTED\_GENERIC\_REFERENCE} occurred in 55 models. These warnings relate to a current limitation in the Ecore parser, where generics are not fully supported. This lack of support leads to element skipping. While the absolute number of skipped elements remains negligible, these warnings highlight where the dataset's heterogeneity interacts with parser and IR abstraction choices, motivating future improvements to the parsing backend.

\textbf{Parsing time and file sizes}. D1.M3 (parsing time) and D1.M4 (file size) are descriptive signals rather than quality scores. They help identify outliers and provide a rough indication of scalability, but should not be interpreted as intrinsic model complexity, as both are influenced by the chosen IR serialization, runtime environment, and implementation details.

Overall, the ArchiMate parser is significantly faster compared to the Ecore parser. Across the three datasets, ArchiMate parsing takes 1.45ms on average (max. 110ms), whereas Ecore parsing averages to 6.9ms (ModelSet) and 17ms (AtlanMod Zoo), with a small number of pronounced outliers reaching 2.5-2.6s. This difference is explained by the processing pipeline, as the ArchiMate parser reads the XML directly and constructs the IR in one pass, while the Ecore parser first loads the \texttt{.ecore} model via \texttt{pyecore} and subsequently maps the instantiated objects into the IR.

File size statistics differ between the two languages. For ArchiMate, the IR is smaller than the source representation, because Archi exports often include diagrammatic metadata that the IR omits. For Ecore, the IR is generally larger than the source. since compact \texttt{.ecore} serializations (without diagram information) expand into a more verbose JSON graph that explicitly materializes nodes, edges, and attributes. The transformation also introduces explicit containment and reference edges (e.g., a \texttt{Package} linked to each contained \texttt{Class}), which further increases the IR size.

\subsection{D2. Lexical Quality}
In the following, we report the measures for the lexical dimension. \autoref{tab:d2-summary} summarizes the main lexical characteristics of the three model datasets under the common configuration (node labels only, name attribute, tokenizer configured as in \autoref{sec:experimental-setup}).

\textbf{Label Presence.} D2.M1 quantifies whether label-eligible nodes carry a non-empty name. Note that all subsequent lexical measures operate on these extracted labels. For both ModelSet and AtlanMod Zoo, label presence is 100\%, which is largely expected, as the Ecore constructs that map to IR nodes all have a mandatory \texttt{name} feature, and typical EMF editors either enforce non-empty names or immediately auto-fill a default value when creating elements. Consequently, D2.M1 provides limited discriminative power for Ecore under the current lexical profile.

For EA ModelSet (ArchiMate), label presence is also high (99.6\%, 382 missing labels), but the distribution is not uniform. On the model level, the median missing share is 0.0, so the typical model is fully labeled. However, the maximum missing share reaches 0.81, indicating that in some models a large fraction of elements have no name. Manual inspection suggests that these are often sketch-like models, consistent with ArchiMate tooling, which often tolerates unnamed elements. This tail of under-labeled models is small in absolute numbers but relevant for tasks that assume that all elements carry meaningful names.

\textbf{Label Length and Single vs Multi-Word Usage.} Label length (D2.M2) and single vs multi-word usage (D2.M3) reveal stylistic differences between the datasets. Because both measures are computed per model and summarized across models, reporting medians is appropriate as lexical distributions are typically heavy-tailed, and medians are less dominated by a small number of verbose labels than means. However, means remain useful for outlier analysis and are already retained in the underlying statistics. In the following, we use the terms ``short label'' for words with fewer than 5 characters or 2 tokens and ``long label'' for words with more than 30 characters or 8 tokens.

For EA ModelSet, the median per-model label length is 16 characters / 2 tokens, with a mean of about 18.1 characters and 2.6 tokens. Furthermore, it has a low median single-word share ($\approx 18\%$), with the median shares of short and long labels at about 18\% and 10\%, respectively. This is consistent with enterprise architecture naming conventions (e.g., ``Business Process'', ``Customer Information'', ``Application Component''), where labels often encode role, function, or scope in multi-word terms. A non-trivial tail of very long labels (per-model medians up to 151 characters / 26 tokens) likely reflects copied fragments from documentation.

In contrast, both Ecore datasets are dominated by short, identifier-like names. ModelSet has a median label length of 7 characters and 1 token (mean 7.6 characters / 1.4 tokens), and a median single-word share of around 57\%. AtlanMod Zoo shows similar behavior, with a median label length of 8.5 characters and 1 token (mean 8.8 characters / 1.4 tokens), and a median single-word share of 55\%. This reflects typical Ecore naming, where many labels are identifiers (e.g., \texttt{name}, \texttt{value}, type names like \texttt{EString}), and multi-word labels are less common.

The ``short/long label shares'' reinforce this interpretation. In Ecore datasets, the median model has a high share of short labels (ModelSet median short-label share is $\approx 0.60$, AtlanMod Zoo is $\approx 0.55$), while long labels are rare (ModelSet mean long-label share is $\approx 0.003$, AtlanMod Zoo is $\approx 0.004$). EA ModelSet shows the opposite pattern with fewer short labels (median $\approx 0.18$) and a noticeably larger fraction of long labels (median $\approx 0.10$). Taken together, these measures show that (i) EA ModelSet exhibits phrasal, multi-word naming with a mix of short and long labels and (ii) ModelSet and AtlanMod Zoo have mainly predominantly short, single-word naming, reflecting their technical and metamodel-centric nature. For NL-facing tasks, this implies that EA ModelSet provides richer surface forms for each element, whereas Ecore datasets provide more regular but less expressive names that may require additional context or normalization. 

\textbf{Lexical Diversity and Vocabulary.}
D2.M4 provides corpus-level vocabulary statistics (total tokens, vocabulary size) and a simple diversity ratio (Type-Token Ratio (TTR), where TTR = vocab / tokens). ModelSet contains the largest token volume (593,637 tokens), but a comparatively small vocabulary (19,366 unique tokens, TTR $\approx 0.13$), indicating heavy repetition of common identifiers and metamodel tokens (e.g., name, value, Type, EString). %\todo{why are these part of the IR node name label?} 
EA ModelSet contains fewer tokens (342,486) but a much larger vocabulary (44,356, TTR $\approx 0.03$), consistent with more descriptive, domain-bearing naming. AtlanMod Zoo is the smallest (50,757 tokens, 6,435 unique tokens, TTR $\approx 0.13$) and therefore naturally exhibits less lexical repetition at the dataset level.

The observed TTR values should be interpreted cautiously because TTR is size-sensitive and typically decreases as corpora grow. Accordingly, the low TTR of ModelSet (0.033) reflects both corpus size and lexical repetitiveness, and it should not be read as a universal ``low-quality'' signal. In the current catalog, these statistics serve as first-order descriptors of lexical richness and repetitiveness. To improve cross-dataset comparability, a follow-up extension could add size-robust diversity measures (e.g., segmental TTR and entropy-based measures). 

From a task-suitability perspective, this means that EA ModelSet and AtlanMod Zoo are lexically richer corpora for natural-language–sensitive experiments (e.g., embedding-based similarity, text-to-model tasks), whereas ModelSet provides a more regular technical vocabulary better suited to studying patterns in metamodel design. More sophisticated diversity measures (e.g., moving-average TTR, MTLD) could further stabilize these comparisons, but the current measures already expose substantial differences.

\begin{table*}[t]
\centering
\caption{Summary of D3 Construct Coverage Metrics}
\label{tab:d3-summary}
\small
\setlength{\tabcolsep}{3pt}
\renewcommand{\arraystretch}{1.15}

\begin{adjustbox}{max width=\textwidth}
\begin{tabularx}{\textwidth}{lP{2.75cm}P{1.75cm}P{2.2cm}P{2.7cm}XXX}
\toprule
\textbf{Dataset} &
\textbf{Constructs in catalogue (n/e)} &
\textbf{Observed constructs} &
\textbf{Dataset coverage (\%)} &
\textbf{Unknown types (nodes / edges)} &
\textbf{D3.M1} &
\textbf{D3.M2} &
\textbf{D3} \\
\midrule
EA ModelSet &
74 (63 / 11) &
74 &
100 &
212 / 1615 &
99.2 &
81.5 &
90.3 \\
ModelSet &
38 (16 / 22) &
38 &
100 &
0 / 0 &
100 &
84.7 &
92.3 \\
AtlanMod Zoo &
38 (16 / 22) &
34 &
89 &
0 / 0 &
89.4 &
83.5 &
86.5 \\
\bottomrule
\end{tabularx}
\end{adjustbox}
\end{table*}

\textbf{Language Usage.} D2.M5 attempts to detect the predominant natural language of labels per model and counts how many models fall into each language. In the EA ModelSet, English is the dominant language, but it only accounts for $\approx 59.7\%$ of models. The remaining $\approx 40\%$ is distributed across at least 25 distinct languages, including Spanish, Portuguese, Russian, German, Dutch, and several others. This reflects that the data originates from different organizations and countries on GitHub. In ModelSet, around 90\% of the models are detected as English, with a long tail over 36 languages. Some of this tail likely corresponds to genuine non-English models (by origin), and some may be misclassifications due to very short or technical labels. AtlanMod Zoo is almost entirely English ($\approx 97\%$) with only a small number of detected languages (6).

Two caveats are important. First, language detection on short, identifier-like labels is noisy. Thus, counts should be interpreted as indicators of lexical heterogeneity rather than as ground truth. Second, multilinguality interacts directly with downstream task design and has practical implications. For English-only NLP or LLM experiments, AtlanMod Zoo is the easiest to use without additional preprocessing. ModelSet is also largely English but would benefit from language filtering. EA ModelSet, by contrast, offers a realistic multilingual modeling scenario that is attractive for multilingual tasks, but for monolingual tasks, it requires explicit language filtering or translation. In this sense, D2.M5 provides actionable evidence about dataset preparation requirements rather than about label ``quality'' per se.

\subsection{D3. Construct Coverage}
The construct coverage dimension D3 operationalizes the extent to which a dataset instantiates the constructs of a modeling language, as defined by a construct catalog. Currently, D3 comprises two measures. D3.M1 (Presence/Coverage) captures whether each catalog construct occurs at least once in the dataset and how much of the catalog a typical model uses. D3.M2 (Frequency/Balance) captures how often each construct is instantiated and how evenly usage is distributed across constructs. The latter is summarized using utilization entropy over the construct-frequency distribution (higher entropy indicates more balanced utilization), and the D3.M2 score is the utilization entropy scaled to $[0,100]$.

\autoref{tab:d3-summary} summarizes the D3 results across the three datasets. Two caveats are important when interpreting these results. First, D3 is catalog-bounded, i.e., ``100\% coverage'' means that all constructs in the catalog occur at least once, not that the dataset covers the full underlying language specification. The ArchiMate catalog used here covers the full set of element and relationship types (from the ArchiMate specification). Additional higher-level concepts, such as viewpoints, are currently out of scope but can be easily added by expanding the catalog. The complete Ecore metamodel is quite complex; thus, the current catalog focuses on core constructs (metaclasses, containment/references, and common cardinality/collection modifiers). Second, D3 depends on the parser–IR–catalog mapping, meaning observed and unknown types reflect the interaction among the dataset, the parser mapping to the IR, and the catalog matching. Consequently, D3 should not be used to claim cross-language ``better coverage'', but to characterize coverage and utilization within the respective catalog.

\textbf{EA ModelSet.} For the EA ModelSet, dataset-level construct presence is complete as all 74 constructs in the ArchiMate catalog (63 node types, 11 edge types) are observed at least once, and all catalog groups (e.g., Business, Application, Technology, Physical, Motivation, Strategy, Implementation/Migration, Relationship, Other) reach 100\% coverage. However, per-model coverage is low to moderate: the median model uses only about 17.6\% of the catalog (§ constructs), with the 75th percentile at around 28\%, and only a few models approach full coverage (max $\approx 0.973$). This pattern is expected for EA models, which are typically organized by viewpoints and scope, resulting in specialized models that focus on a subset of the modeled architecture (e.g., the business layer or the technology layer) and, consequently, use only a subset of the language to represent it.

% D1.M1. Score = 100.0 * coverage_share_dataset * (1.0 - unknown_type_share_dataset)
% score = max(0.0, min(100.0, score))

\begin{figure*}
    \centering
\includegraphics[width=\linewidth]{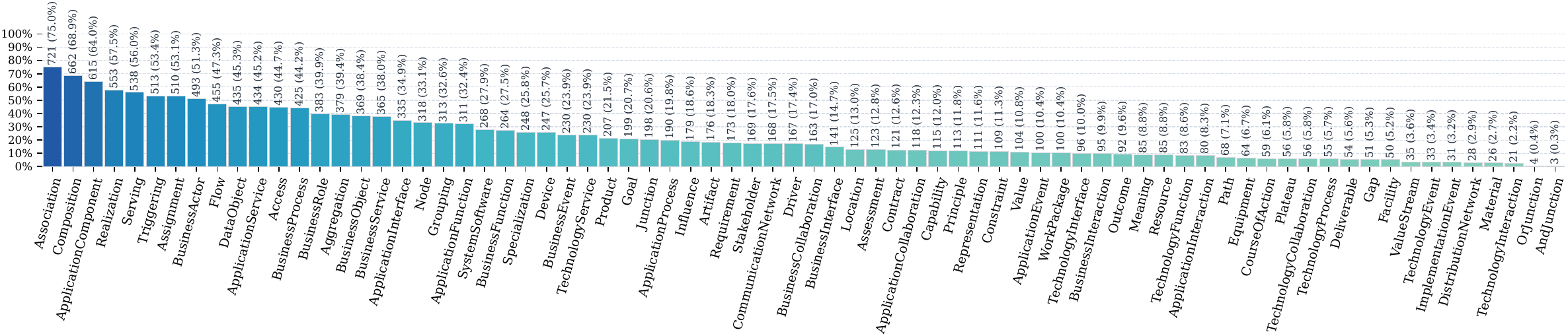}
    \caption{EA ModelSet Construct Coverage}
    \label{fig:ea-modelset-construct-coverage}
\end{figure*}

EA ModelSet is also the only dataset in our study with a nonzero share of unknown types ($\approx 0.008$), corresponding to 212 unknown node types and 1615 unknown edge types. The most frequent unknown examples (e.g., \texttt{Realisation}/\texttt{Specialisation},  legacy \texttt{UsedBy}, and \texttt{Infrastructure*} types) are consistent with tool- or version-induced naming variants and normalization gaps rather than entirely novel constructs. These are mostly spelling issues (e.g., Realisation vs Realization, Specialisation vs Specialization) as some tools represent these differently or constructs from older ArchiMate versions (e.g., older Infrastructure* vs their newer Technology counterparts or UsedBy renamed to Serving in ArchiMate 3.0). The parser can be configured to normalize such variants, but in the reported configuration, these variants were intentionally preserved to surface standardization and normalization issues. D3.M1 thus exposes both the breadth of construct usage and small but relevant mismatches between the mined models, the parser, and the catalog.

In terms of frequency, EA ModelSet contains 228,028 construct instances. As expected, usage is dominated by structural relationships such as Composition ($\approx 25k$ instances), Association ($\approx 20k$), Realization ($\approx 15k$), and Serving ($\approx 12k$), alongside frequently used node types like BusinessProcess ($\approx 9k$) and ApplicationComponent ($\approx 8.8k$). The utilization entropy is about 0.815, yielding a D3.M2 score of 81.5. This distribution is more skewed than the Ecore datasets and reflects that a small set of relationship types (e.g., containment and association) is pervasive, while many constructs (e.g., motivation/strategy elements) are comparatively rarely used. This skew must be accounted for in downstream tasks, e.g., when creating test/train splits. \autoref{fig:ea-modelset-construct-coverage} shows an overview of the construct coverage in the EA ModelSet.

\begin{figure*}[ht]
    \centering
    \includegraphics[width=.7\linewidth]{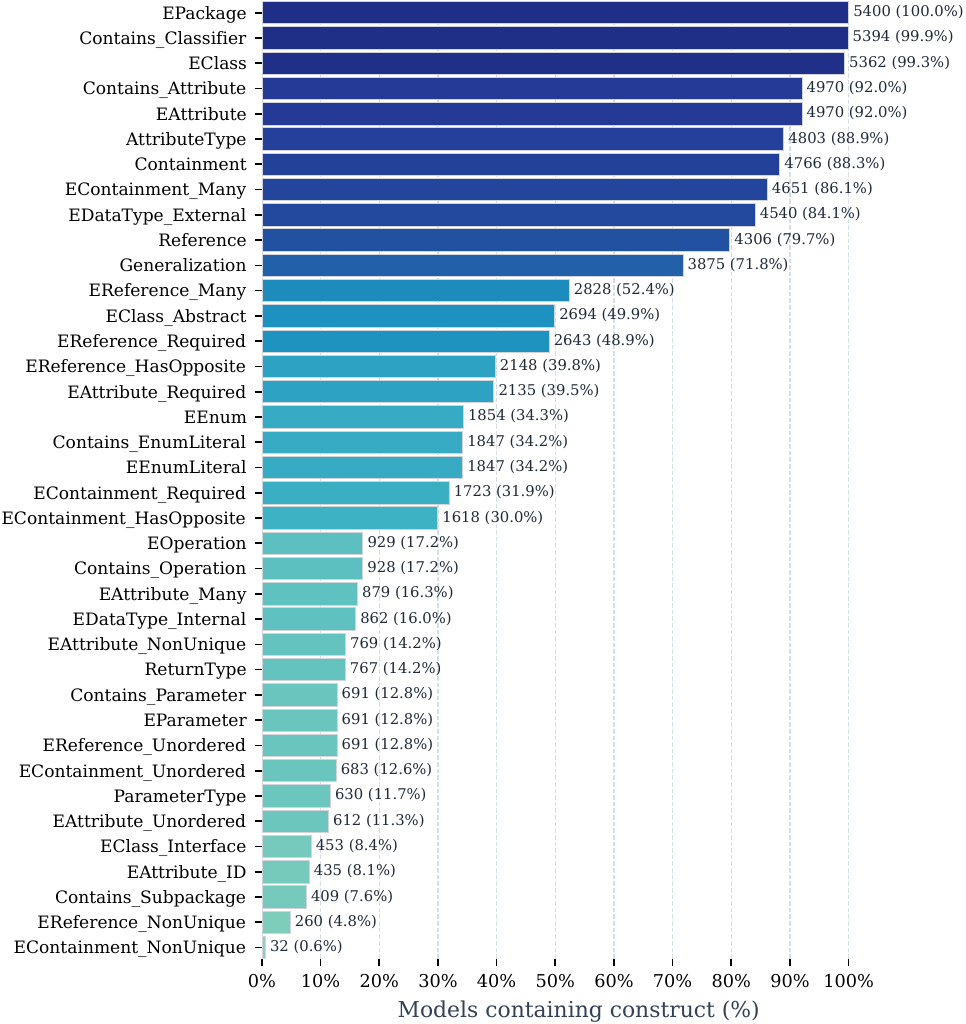}
    \caption{ModelSet Construct Coverage}
    \label{fig:modelset-construct-coverage}
\end{figure*}

\textbf{ModelSet.} ModelSet achieves full catalog coverage under the current Ecore construct catalog (38/38 observed), with zero unknown types and full group-level coverage, leading to a D3.M1 score of 100. Per-model coverage is substantially higher than for EA ModelSet. The median model uses around 39.5\% of the catalog ($\approx 15/38$ constructs with mean $\approx 41.5\%$, 75th percentile $\approx 52.6\%$, maximum $\approx 92.1\%$). In other words, a typical metamodel uses roughly 40\% of the core Ecore constructs, and a subset of models comes close to using almost the entire catalog, suggesting that many metamodels exploit a broad slice of the Ecore feature set.

\begin{table*}[t]
\centering
\caption{Summary of D4 Size Metrics}
\label{tab:d4-summary}
\scriptsize
\setlength{\tabcolsep}{3pt}
\renewcommand{\arraystretch}{1.15}

\begin{adjustbox}{max width=\textwidth}
\begin{tabularx}{\textwidth}{@{}l
>{\raggedright\arraybackslash}p{2.45cm}
>{\raggedright\arraybackslash}p{3.6cm}
>{\raggedright\arraybackslash}p{2.4cm}
>{\raggedright\arraybackslash}p{2.2cm}
>{\raggedright\arraybackslash}X@{}}
\toprule
\textbf{Dataset} &
\makecell[l]{\textbf{Elements}\\\textbf{(nodes / edges)}} &
\makecell[l]{\textbf{Per-model elements}\\\textbf{(median/mean/P75/max)}} &
\makecell[l]{\textbf{Per-model degree}\\\textbf{(median/mean)}} &
\makecell[l]{\textbf{Isolated nodes}\\\textbf{(median/mean)}} &
\makecell[l]{\textbf{Components per model}\\\textbf{(median/mean/P75/max)}} \\
\midrule
EA ModelSet &
\makecell[l]{229,855\\(103,334 / 126,521)} &
111 / 239 / 222 / 8554 &
2.36 / 2.39 &
4 / 21.3 &
5 / 25.42 / 16 / 3938 \\
ModelSet &
\makecell[l]{1,009,192\\(331,610 / 677,582)} &
62 / 186.89 / 182 / 3434 &
3.63 / 3.62 &
0 / 0.016 &
1 / 1.01 / 1 / 29 \\
AtlanMod Zoo &
\makecell[l]{88,832\\(27,305 / 61,527)} &
106 / 296.1 / 281.5 / 8779 &
4.28 / 4.36 &
0 / 0.003 &
1 / 1.09 / 1 / 2 \\
\bottomrule
\end{tabularx}
\end{adjustbox}
\end{table*}

Construct frequencies confirm this picture. With approximately 1.24 million construct instances, ModelSet is by far the largest corpus. Central metamodel constructs appear with high counts (EClass $\approx 141k$, Contains\_Classifier $\approx 150k$, Generalization $\approx 116k$, Reference and Containment around 58–84 k), and cardinality and collection modifiers (Required, Many, Unordered, NonUnique) are well represented. Utilization entropy is the highest among the three datasets (entropy $\approx 0.847$, D3.M2 score $\approx 84.7$), suggesting comparatively more balanced usage. This makes ModelSet a strong candidate for evaluating approaches that rely on diverse, numerous examples of Ecore constructs, such as automatic metamodel completion or pattern mining. \autoref{fig:modelset-construct-coverage} shows an overview of the construct coverage in the ModelSet model dataset.

\textbf{AtlanMod Zoo.} AtlanMod Zoo uses the same Ecore construct catalog as ModelSet, but does not achieve full construct coverage. Of the 38 constructs, 34 are observed, with no unknown types, yielding a D3.M1 score of 89.4. The missing constructs are part of two groups: (EAttribute\_ID, EClass\_Interface) and (EContainment\-\_NonUnique, EReference\_NonUnique). Thus, group-level coverage remains 100\% for metaclasses, containment, typing, references, and cardinality, but drops for modifier (1/3) and collection (4/6). This aligns with a typical property of curated datasets as curation improves consistency and reduces noise, but does not imply maximal diversity of specialized language features.

Per-model coverage is nevertheless high, with the median model covering about 52.6\% of the catalog ($\approx 20/38$ constructs), and the upper 75th percentile around 57.9\%. The dataset contains roughly 136k construct instances, and the frequency distribution again shows a strong presence of EClass, Generalization, Reference, and Containment. Collection-related modifiers such as EAttribute\_Unordered and EAttribute\_NonUnique are very frequent in this dataset, whereas operations and parameters are almost absent (EOperation and EParameter occur only a few dozen times). Utilization entropy is high (entropy $\approx 0.835$, D3.M2 score $\approx 83.5$), slightly below that of the ModelSet but above that of the EA ModelSet. Overall, AtlanMod Zoo is a high-quality but structurally narrower Ecore corpus, which helps explain its lower D3 score despite curation and is consistent with its role as a manually curated collection of ``didactic'' metamodels. 

\subsection{D4. Size}

Dimension D4 provides descriptive signals about the scale and structural shape of the parsed models in the IR. These measures are not intended to express ``good'' or ``bad'' quality, but they characterize the operational properties of the parsed graphs and are useful for (i) understanding dataset structure, (ii) identifying outliers and potential parsing anomalies, and (iii) anticipating scalability constraints for downstream analyses. Dimension D4 currently consists of four measures: D4.M1 describes model size (nodes, edges, total elements, edge–node ratio), D4.M2 describes average degree (how many relationships a node participates in), D4.M3 captures connectivity (number of connected components, largest component, isolated nodes), and D4.M4 characterizes containment depth (depth of containment trees and the number/share of nodes that are contained vs. roots). Because all measures are computed over the IR, results depend on IR design and parser-mapping choices (e.g., which relationships become edges, whether implicit links are made explicit) and on language-specific semantics (especially for containment). Cross-language comparisons must be qualified in particular, since Ecore graphs tend to include many structural relations that are implicit in the metamodel semantics (e.g., packages containing classes, classes containing attributes), whereas ArchiMate graphs primarily represent explicit relationships modeled by the author.

\autoref{tab:d4-summary} summarizes the main D4 metrics for the three datasets. In terms of overall corpus size, ModelSet is by far the largest dataset, with 1,009,192 IR elements (331,610 nodes and 677,582 edges), followed by EA ModelSet with 229,855 elements (103,334 nodes and 126,521 edges), and AtlanMod Zoo with 88,832 elements (27,305 nodes and 61,527 edges). At the per-model level, the distributions are strongly heavy-tailed across all datasets, as reflected by the gap between the median and mean values and by large maxima. In the EA ModelSet, a typical model contains 111 elements (median), while the mean rises to 239 due to a small number of very large models (maximum 8,554). ModelSet exhibits smaller typical models (62 median elements) but still a substantial tail (mean 186.89, maximum 3,434). AtlanMod Zoo contains fewer models but larger ones on average; the median is 106 elements, and the mean increases to 296.1, with a maximum of 8,779. This indicates that a small number of extremely large metamodels are present in the AtlanMod Zoo, despite the corpus being curated.

The degree and edge–node ratio statistics highlight clear differences in graph density. EA ModelSet has the lowest degree values (median 2.36, mean 2.39), indicating comparatively sparse graphs in which nodes participate in only a few relationships on average. Both Ecore datasets are denser, consistent with the Ecore IR materializing many structural relations as edges. ModelSet has a median degree of 3.63 (mean 3.62), while AtlanMod Zoo is the densest, with a median degree of 4.28 (mean 4.36). These density differences are operationally relevant because they directly affect the cost of graph-based analyses, such as graph algorithms and ML models.

Connectivity (D4.M3) is the strongest differentiator between the EA and Ecore corpora. EA ModelSet contains a non-negligible number of isolated nodes (median 4, mean 21.3) and highly fragmented graphs, as the median number of components per model is 5, but the mean increases to 25.42, and the maximum reaches 3,938. The isolated-node share can reach 100\% in extreme cases, i.e., there are models in which no node is connected to any other. This indicates that many ArchiMate models consist of documented EA Smells~\cite{SmajevicEtal21KG4EASmells}, in this case multiple disconnected fragments, and that some models include large quantities of unconnected elements, which is plausible for mined EA artifacts, where elements may be created without establishing explicit relations. Also, this is consistent with the behavior of common EA tools, which allow modelers to create nodes that never get wired into the main diagram, and some tools do not actively clean up such dangling elements or encode certain ``visual'' groupings as explicit edges. The resulting graphs are realistic but structurally messy from a pure graph perspective. In contrast, both Ecore model datasets are almost always connected. ModelSet has essentially no isolated nodes (median 0, mean 0.016) and is dominated by single-component graphs (median component size 1, mean 1.01, maximum 29). AtlanMod Zoo shows the same pattern even more strongly (isolated nodes: median 0, mean 0.003; components: median 1, mean 1.09; maximum 2), reflecting that curated metamodel examples typically form a single coherent containment-structured graph. For any algorithm that assumes connected graphs, ModelSet and AtlanMod Zoo are therefore ``clean'' corpora, while EA ModelSet is a realistic test of robustness to disconnected and noisy structures.

Containment depth (D4.M4) adds a hierarchical perspective, with an important caveat: the computation approximates the longest containment paths via a BFS-style propagation capped by the number of nodes, so extreme maximum depths often signal pathological or cyclic patterns rather than meaningful hierarchical designs. Containment semantics are also language-specific: in ArchiMate, composition and aggregation edges are the only containment relationships, whereas Ecore uses explicit containment relations and containment=true flags. Thus, cross-language comparisons of depth must be read with caution. With that in mind, EA ModelSet exhibits very shallow containment, as for typical models, the median maximum depth is 1 (75th percentile 2), and the median mean depth per model is around 0.23, with only about 20\% of nodes contained (median contained-node share) and many elements at the root level. Yet a few models show extreme maximum depths (up to 2,880), most likely due to long chains of nested composite elements. In ModelSet, containment is much more central, with a median maximum depth of 4 (p75 6), a median mean depth of around 1.8, and a median contained-node share of approximately 86\%, indicating that most nodes sit somewhere in a hierarchical structure below a small set of roots. AtlanMod Zoo is similar in this regard as the median maximum depth is 4 (p75 $\approx 6.25$), median mean depth around 1.68, and the median contained-node share is about 94\%, with very few root nodes. Both Ecore datasets, therefore, represent metamodels as deeply hierarchical graphs, whereas ArchiMate models in EA ModelSet are predominantly flat, with only a few extremely deep hierarchies.

Taken together, the D4 results show that the datasets occupy distinct structural regimes. EA ModelSet consists of many medium-sized models plus a few very large ones, with relatively sparse, fragmented, and mostly shallow graphs that closely mirror real-world EA artifacts, including their ``messiness'' (as expected from mined data). ModelSet is a broad corpus of small-to-medium metamodels that are denser, almost always connected, and strongly hierarchical. Finally, AtlanMod Zoo is a smaller but curated set of comparatively large, dense, and strongly hierarchical metamodels.

\subsection{Cross-Dimensional Summary}\label{sec:cross-dimensional-summary}

\ins{Across D1--D4, the benchmark results show that the three datasets are not simply ``better'' or ``worse'' versions of one another, but represent different dataset profiles. All three corpora are technically usable with the selected parsers, with high parseability and negligible information loss due to skipped elements. The main differences concern the kinds of evidence each dataset provides in lexical richness and multilinguality, construct breadth and balance, and graph structure. This supports the benchmark's role as a decision aid for dataset selection, rather than as a single quality ranking.}

\ins{EA ModelSet is the most realistic but also the most heterogeneous corpus in the study. It combines high parseability with descriptive, often multi-word labels, broad ArchiMate construct coverage, and substantial multilinguality. At the same time, its models are structurally sparse, fragmented, and sometimes noisy, with many disconnected components and a small number of extreme outliers. This makes EA ModelSet useful for evaluating methods that should tolerate heterogeneous naming practices and incomplete or disconnected structures. For monolingual NLP or graph algorithms that assume connected inputs, however, explicit preprocessing such as language filtering or outlier handling is required.}

\ins{ModelSet's main strength is scale as it contains the largest number of models and IR elements, reaches full coverage of the current Ecore construct catalog, and shows the most balanced construct utilization among the three datasets. Its labels are mostly short, identifier-like, and repetitive, which makes it less expressive for natural-language-heavy experiments but well suited for metamodel mining, completion, classification, and other tasks that benefit from many structurally comparable Ecore examples. The warnings also indicate that mined Ecore corpora exhibit parser and dependency issues that should be explicitly reported when the dataset is used in experiments.}

\ins{AtlanMod Zoo provides the cleanest and most curated profile. It is smaller than ModelSet, but its models are highly parseable, predominantly English, dense, connected, and strongly hierarchical. Its lower construct-coverage score reflects a narrower selection of Ecore features and does not indicate poor quality. Consequently, AtlanMod Zoo is useful as a compact, controlled Ecore corpus, while ModelSet is more appropriate when scale and construct diversity are required.}

\ins{Overall, the benchmark dimensions help identify which dataset properties matter for a given downstream experiment and which preprocessing steps should be made explicit for reproducibility. The evaluation suggests three practical selection rules. Researchers needing realistic, multilingual, and structurally messy EA data should prefer EA ModelSet. Researchers who need large-scale, construct-diverse Ecore data should prefer ModelSet. Researchers who need a small, clean, and mostly English Ecore corpus should prefer the AtlanMod Zoo. The benchmark dimensions help identify which dataset properties matter for a given downstream experiment and which preprocessing steps should be made explicit for reproducibility.}

\section{Discussion}\label{sec:discussion}
This section synthesizes the main contributions and findings of our research and reflects on their implications for research and practice, as well as their limitations and threats to validity.

\subsection{Implications for Research and Practice}
Our results demonstrate that the platform can reliably parse heterogeneous corpora and surface descriptive signals that are directly actionable for dataset curation and task design. First, parsing outcomes (D1) indicate high technical usability across all three datasets under a common configuration, with negligible information loss from skipped elements and clear diagnostics that point to tool- or dialect-specific idiosyncrasies. This suggests that benchmarking can be used early in the dataset lifecycle to quantify improvements from normalization and filtering, track representation changes across language and tool versions, and make explicit inclusion and deduplication decisions that mitigate bias.

Second, lexical signals (D2) reveal big differences in label regimes across datasets. Ecore-based corpora exhibit near-complete label presence due to mandatory naming features in typical modeling editors, reducing discriminative power for presence metrics under the current profile, whereas the EA ModelSet models contain a realistic tail of under-labeled sketches that is small in absolute terms but relevant for text-based tasks. Multilinguality further varies: AtlanMod Zoo is predominantly English, ModelSet is largely English with a long tail, and EA ModelSet is genuinely multilingual, which directly impacts preprocessing pipelines for monolingual versus multilingual NLP/LLM tasks. Consequently, D2 measures offer concrete guidance for model dataset preparation (e.g., language filtering, label completion) rather than a singular notion of label “quality”.

Third, structural substance (D4) shows that the model datasets occupy distinct regimes: Ecore datasets are dense, connected, and strongly hierarchical graphs. In contrast, the ArchiMate models in the EA ModelSet are predominantly flat, sparse, and fragmented with a few deeper hierarchies. This contrast mirrors differences in metamodel conformance and modeling practice and is essential when aligning datasets to downstream tasks that rely on structural variation (e.g., graph-based classification, completion, or refactoring). 

Taken together, these observations reinforce the idea that benchmarking should make salient characteristics explicit—parseability, label regimes, construct coverage, and structural distributions—so researchers can assess task suitability and improve comparability across studies.

Beyond these scientific insights, the framework and platform enable several practical implications as well:
\begin{itemize}
    \item Systematic analysis of existing model datasets at scale, beyond the initial corpora reported here, including quantifying variance introduced by parser mappings and configuration choices.
    
    \item Synthetic dataset generation guided by observed distributions (e.g., size, depth, connectivity), supporting controlled experiments that probe boundary conditions of ML pipelines and transformation frameworks.

    \item Development and evaluation of a coherent taxonomy of model dataset metrics and their interpretation, building on established software and model quality research while preserving language-agnostic comparability through the IR.

    \item Standardized reporting of dataset properties as part of experimental sections, improving transparency, reproducibility, and cross-study comparison, especially when models originate from mined, curated, or synthetic sources with different noise patterns.
\end{itemize}

Finally, reproducibility and task alignment benefit from explicit benchmark profiles. Profiles capture parser selection, lexical tokenization, and enabled measures, and thus act as reproducibility units for repeated runs under identical settings, while making configuration-dependent outcomes explicit in benchmark evidence.

\subsection{Limitations}\label{sec:limitations}
While the framework advances systematic dataset characterization, several limitations currently remain. 

First, parser robustness and measuring outcomes are conditional on the current mappings into the graph-based IR. As a result, reported values reflect the implemented parser semantics and profile configuration, not a universal ground truth across all serializations and tool dialects. \ins{This affects almost all benchmarking measures as they are based on the IRs that the selected parser constructs. Different parsers for the same language may expose different levels of detail. A metamodel-based parser may produce a more verbose graph, while a handcrafted parser may omit unsupported concepts or intentionally abstract from low-level details. Consequently, cross-dataset comparisons are most reliable when datasets are processed with the same parser version, benchmark profile, and construct catalog. When different parsers or profiles are used, benchmark results should be interpreted as conditional on these mappings and read together with parser diagnostics such as warnings, skipped elements, unknown types, and IR size.} Extending parser completeness and portability to handle language versions and tool-specific formats uniformly is ongoing work.

Second, edge label eligibility is a known challenge. For Ecore, many IR edges reflect implicit containment without meaningful names; naïvely including them artificially increases label presence. A natural extension is to define eligibility by relation type (e.g., constrain to EReference edges) to avoid conflating structural scaffolding with semantically meaningful labels.

Third, the current lexical profile focuses only on node-name attributes, using a shared tokenizer. This simplifies comparability but limits coverage of other potentially informative textual features (e.g., documentation, stereotypes, tagged values) that vary across languages and tools.

Fourth, the present study focuses on introducing the framework and platform rather than establishing causal links between benchmarked measures and downstream ML task performance. A structured evaluation of how descriptive measures (D1–D4) relate to specific tasks (e.g., domain classification, partial model completion) requires dedicated systematic research. \ins{Such task-oriented evaluations may also depend on the availability and quality of annotation layers.}

\ins{Annotation benchmarking can be integrated into future versions of the framework as an additional quality dimension. The existing conceptual model already supports this extension through optional \texttt{label?} attributes on \texttt{ModelArtifact} and \texttt{ModelElement}, together with new measures and metrics that can operate at dataset, model, or element level. Model-level labels can capture categories such as domains or task classes, while element-level labels can identify annotated elements or subgraphs by assigning the same label to multiple model elements. Annotation files can also be represented as \textit{SupplementaryArtifacts} and linked to the corresponding models or model elements through stable identifiers and machine-readable mappings. Measures could then assess annotation coverage, completeness, consistency, class balance, inter-rater agreement, provenance, and conformity with an explicit annotation scheme.}

Finally, not all practically relevant aspects are amen\-able to uniform benchmarking. Licensing, redistribution, provenance, and confidentiality constraints shape dataset usability but require manual assessment and cannot be captured fully by automated measures.

\subsection{Threats to Validity}
Several threats may affect the interpretation of our findings. 

\paragraph{Construct validity. }
% Construct validity concerns whether the operationalizations used in a study accurately represent the theoretical concepts they are intended to measure. Are we truly measuring what we claim to measure?
Configuration dependence is central: parser selection, IR mapping, and lexical profile settings determine which elements are materialized and how labels are tokenized, potentially shifting construct coverage, size distributions, and lexical statistics in ways that reflect configurations more than underlying datasets. 
Heterogeneity in serialization and tool dialects remains a confounder, as different formats preserve distinct information layers (e.g., diagram layout vs. abstract syntax), making cross-dataset comparisons sensitive to representation choices.
Additionally, language detection on short, identifier-like labels is noisy and should be treated as indicative of lexical heterogeneity rather than as ground truth, especially in monolingual NLP setups.

\paragraph{External validity. }
The origins of model datasets further introduce threats. Mined corpora tend to be noisier, including sketches and partial models; curated sets are cleaner but may reflect artificial or simplified scenarios; industrial artifacts are larger and validated but rarer and sometimes subject to confidentiality constraints.

\ins{A related external validity threat concerns the generalizability of the platform requirements themselves. The platform requirements were derived from the literature, stakeholder goals, and our experience with selected model datasets, but they have not yet been validated through a systematic requirements study with a broader community of dataset providers and benchmark users. Consequently, the current requirements may underrepresent needs arising in other datasets, modeling languages, or downstream tasks. We mitigate this by treating the requirement set and measure catalog as an initial version and by designing the platform around extensible parsers, measures, profiles, and reporting artifacts.}

\ins{A further threat concerns the diversity of the datasets analyzed in this study. The evaluation demonstrates the framework's applicability and the platform's extensibility across multiple datasets and modeling languages. However, the analyzed datasets are predominantly composed of structural models. Consequently, the evaluation does not yet provide evidence that the framework and platform generalize equally well to behavioral, executable, simulation-oriented, or other domain-specific models, which may require additional language-specific parsers and measures. In the future, we will expand our platform to further modeling languages, dimensions, measures, and metrics to further strengthen its potential value.}

%A further threat concerns the diversity of the datasets analyzed in this study. Although they conform to different modeling languages, they are predominantly composed of structural models. Consequently, the evaluation does not yet provide evidence that the framework and platform generalize equally well to behavioral, executable, simulation-oriented, or other domain-specific models, which may require additional language-specific parsing and measurement capabilities.}

\del{From a generalization perspective, we showed in this paper that our framework is applicable and the platform is extensible to multiple model datasets, conforming to heterogeneous modeling languages. In the future, we will expand our platform to further modeling languages, dimensions, measures, and metrics to further strengthen its potential value.}

\paragraph{Internal validity. }
% Internal validity concerns whether observed effects or differences can be attributed to the factors under investigation rather than to confounding variables or design flaws.
Duplicates and near-duplicates within datasets present another threat. If identical or highly similar models are included multiple times, observed effects may be driven by redundancy rather than genuine structural or lexical properties of the corpus, inflating statistics or violating independence assumptions in empirical evaluations.

\subsection{Roadmap}\label{sec:future-work}

\ins{The limitations above open several directions for future research and community extensions of the benchmarking framework and application. We intend to evolve the benchmarking framework and application along the following key pillars:}

\paragraph{\ins{Broader datasets, languages, and formats.}}
\ins{The present evaluation focuses on ArchiMate and Ecore datasets and therefore only covers "structural" models. A first direction is to extend the platform to further modeling languages, serialization formats (e.g., additional UML and ArchiMate formats), modalities (e.g., tables, images), and use cases, including behavioral, executable, simulation-oriented, or domain-specific models. Candidate extensions include BPMN and DMN process models~\cite{saeedi2025empirical}, Petri nets~\cite{hillah2017petri}, Simulink models~\cite{BollBAKV21}, as well as larger multi-notation datasets~\cite{sola2022signavio}. This would strengthen the empirical claims about general applicability and would also stress-test the IR against languages whose semantics are behavioral rather than structural. In the same direction, the current Ecore support should be extended from \texttt{.ecore} metamodel files to instance models conforming to Ecore-based metamodels.}

\paragraph{\ins{Language-specific and domain-specific measures.}}
\ins{While the framework deliberately emphasizes language-agnostic measures over the shared IR, future versions should add more language-specific measures where this improves interpretability. The current construct coverage dimension (D3) already provides one such extension point, because construct catalogs encode language-specific concepts while still producing comparable coverage and frequency evidence. Future measures could capture language-specific modeling patterns, constraint violations, well-formedness properties, semantic dependencies, or known smells that cannot be interpreted from generic graph structure alone. For example, Ecore-specific measures could target metamodel design smells such as overly deep inheritance hierarchies or missing opposite references, UML-specific measures could consider diagram-type coverage or class-association balance, and ArchiMate-specific measures could consider layer coverage or viewpoint usage. Such measures would complement the language-agnostic core by making domain semantics explicit when necessary to assess dataset suitability. More generally, the measure catalog can be better aligned with established model quality dimensions in the literature (e.g., consistency, maintainability, complexity, understandability, redundancy, bias).}

\paragraph{\ins{Richer lexical and supplementary evidence.}}
\ins{The current lexical dimension is intentionally narrow and focuses on node names and optionally edge names. Future work should broaden textual feature extraction to documentation fields, comments, stereotypes, tagged values, annotations, constraints, diagram labels, and other supplementary artifacts where available. This is particularly relevant for NLP- and LLM-oriented tasks, where natural-language context may reside outside element names. Such extensions will require profile-level control over which textual sources are included, because different languages and tools expose different textual layers, and because indiscriminately including all text may confound model-intrinsic naming quality with external documentation quality.}

\paragraph{\ins{Annotation benchmarking.}}
\ins{Many downstream tasks require labels or other annotations in addition to the models themselves. Future work should therefore treat annotations as first-class benchmark targets. Measures could assess annotation coverage, schema conformity, class balance, inter-annotator agreement, provenance, granularity, and traceability from annotations to models or individual model elements. A challenge here is the heterogeneity of annotation representations and attachment mechanisms, as they may be stored in separate files (and linked through external identifiers), embedded in custom serialization formats, or maintained in databases. The framework could address this through annotation-specific adapters that normalize these sources into a common annotation representation.}

\paragraph{\ins{Relating benchmark evidence to downstream tasks.}}
\ins{The present framework characterizes datasets descriptively, but does not yet establish systematic relationships between benchmark dimensions and task outcomes. Future studies should connect measures and metrics to concrete tasks such as model classification, clustering, retrieval, completion, repair, transformation, or code generation. For example, lexical diversity and language usage could be related to embedding-based retrieval performance, construct coverage to completion performance on rare constructs, and structural connectivity to graph-learning tasks. Such studies would help distinguish generally informative dataset descriptors from measures that are only relevant under specific task assumptions.}

\paragraph{\ins{Scalability and robust execution.}}
\ins{Larger datasets introduce new technical and methodological challenges. Scaling to hundreds of thousands of models (e.g., as in the SAP-SAM Signavio dataset~\cite{sola2022signavio}) will require more efficient parsing, incremental execution, caching, streaming aggregation, parallel measurement, and report generation that remains usable under large result sets. Scalability work should also include robustness to heterogeneous serialization variations, missing dependencies, corrupted files, and partial representations, because large mined datasets typically contain all of these cases.}

\paragraph{\ins{Benchmark profile usability.}}
\ins{Finally, benchmark profiles are central for reproducibility but currently require users to edit JSON configuration files directly. Future work should improve usability through schema validation, clear diagnostics for invalid values, reusable profile templates, and possibly a small DSL or graphical profile editor. A profile editor could guide users through parser selection, dataset path resolution, measure selection, and parameter validation. This would lower the barrier for adoption without weakening reproducibility.}

\paragraph{\ins{Standardized benchmark reporting.}}
\ins{To improve comparability and reproducibility across empirical studies, future work should define a standardized reporting format for benchmark results. Such reports could accompany scientific publications as appendices or supplementary material and include the benchmark profile, metric summaries, construct distributions, and a concise characterization of the evaluated datasets. Establishing a common reporting convention would facilitate cross-study comparisons and support the accumulation of benchmarking evidence across the conceptual modeling community.}

\paragraph{\ins{Dataset lifecycle benchmarking.}}
\ins{Integration of benchmarks into the dataset lifecycle to quantify deltas across curation phases (raw $\rightarrow$ processed $\rightarrow$ annotated), including shifts in parseability, construct distributions, duplication, and language heterogeneity that impact evaluation reliability.}

\section{Conclusion}\label{sec:conclusion}
Model-driven engineering increasingly depends on model datasets for training and evaluating data-driven and LLM-enhanced techniques, yet such datasets are often created ad hoc and lack standardized, comparable characterizations, hindering reproducibility and interpretability of results. 

To address this gap, this paper introduced a benchmarking framework for model datasets, comprising a metamodel that underpins reproducible, traceable, and extensible assessments, and an initial catalog of quality dimensions and measures that capture parsing robustness, lexical signals, construct coverage, and structural substance. We further implemented a platform that normalizes models into a typed, graph-based intermediate representation and executes a file-based, profile-driven pipeline (Scan → Parse → Measure → Report) with persisted artifacts for inspection and reuse. Currently, our platform supports UML, ArchiMate, and Ecore, enabling cross-language evidence collection under a shared intermediate representation.

We invite the community to adopt the platform to characterize their datasets and to include the generated reports in publications to strengthen comparability and develop higher-quality synthetic datasets based on benchmarking metrics. The platform is publicly available as a Github repository~\cite{BenchmarkingPlatformRepo}, and we welcome community contributions like additional parsers, construct catalogs, and additional measures to broaden coverage and interpretability. A short screencast demonstrating the platform in use is available at \url{https://youtu.be/dxIDuV8eo3U}. A replication package of the benchmarking application and the input datasets used in the evaluation is available at \url{https://doi.org/10.5281/zenodo.21480742}.

Finally, the roadmap presented in \autoref{sec:future-work} outlines several directions for extending the proposed benchmarking framework. %, including the expansion of quality metrics, support for additional modeling languages, standardized reporting practices, empirical validation with downstream ML/LLM tasks, and integration across the dataset lifecycle. These efforts will further strengthen the framework's applicability, reproducibility, and value for conceptual modeling research.

\begin{acknowledgements}
This study was partially funded by the FFG-funded projects Smart GLSP -- Facilitating Large Language Models for Smart GLSP-based Modeling (Grant number: FO999925707) and EAGLE -- Enterprise Architecture Knowledge Graph for Learning and Exploration (Grant number: FO999925702).
This project has been further funded by the Spanish Ministry of Science, Innovation, and Universities under contracts PID2021-125527NB-I00, PID2025-172454NB-I00 and Universidad de Malaga under project JA.B1-17 PPRO-B1-2023-037.
\end{acknowledgements}

% \begin{acknowledgements}
%   We would like to thank all authors who submitted papers.
%   We would also like to thank our reviewers for their efforts and high-quality reviews, which greatly contributed to improving the selected papers.
%   Finally, we would like to express our gratitude to the SoSyM editorial office, specifically to Martin Schindler and Bernhard Rumpe who were always very helpful and supportive.
% \end{acknowledgements}

\section*{Biographies}

\noindent
\begin{minipage}{\columnwidth}
\begin{wrapfigure}{L}{0.3\columnwidth}
  \vspace{-20pt}
  \centering
  \includegraphics[width=0.25\columnwidth]{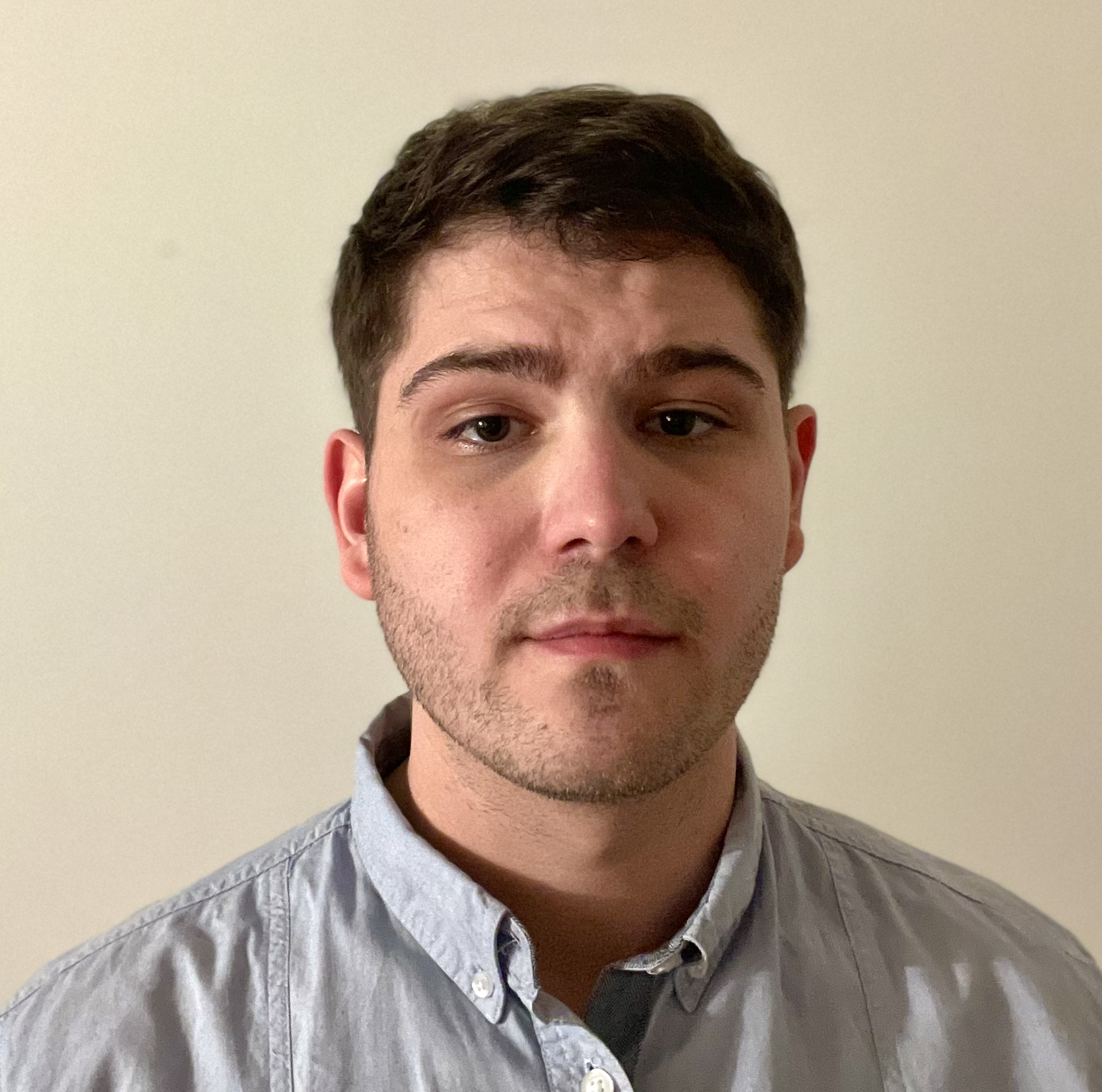}\\
  \vspace{-10pt}
\end{wrapfigure}
{\raggedright
\textbf{Philipp-Lorenz Glaser}\\
	TU Wien \\
	Business Informatics Group\\
	Erzherzog-Johann-Platz 1, 1040, Vienna, Austria\\
	\textbf{philipp-lorenz.glaser@tuwien.ac.at}
\\
~
\\
}
\end{minipage}

\noindent Philipp-Lorenz Glaser is a PhD student at the Faculty of Informatics at TU Wien and a member of the Business Informatics Group. His research interests lie at the intersection of conceptual modeling, software engineering, and artificial intelligence. His current work focuses on the creation, curation, and analysis of model datasets, as well as on assessing their quality and suitability for data-driven modeling research. Further interests include knowledge graphs and ontologies, AI-supported modeling using machine learning and large language models, and the development of modeling languages and tools. For more information, you can contact the author at \url{philipp-lorenz.glaser@tuwien.ac.at}.

\vspace{10px}

\noindent
\begin{minipage}{\columnwidth}
\begin{wrapfigure}{L}{0.3\columnwidth}
  \vspace{-20pt}
  \centering
  \includegraphics[width=0.25\columnwidth]{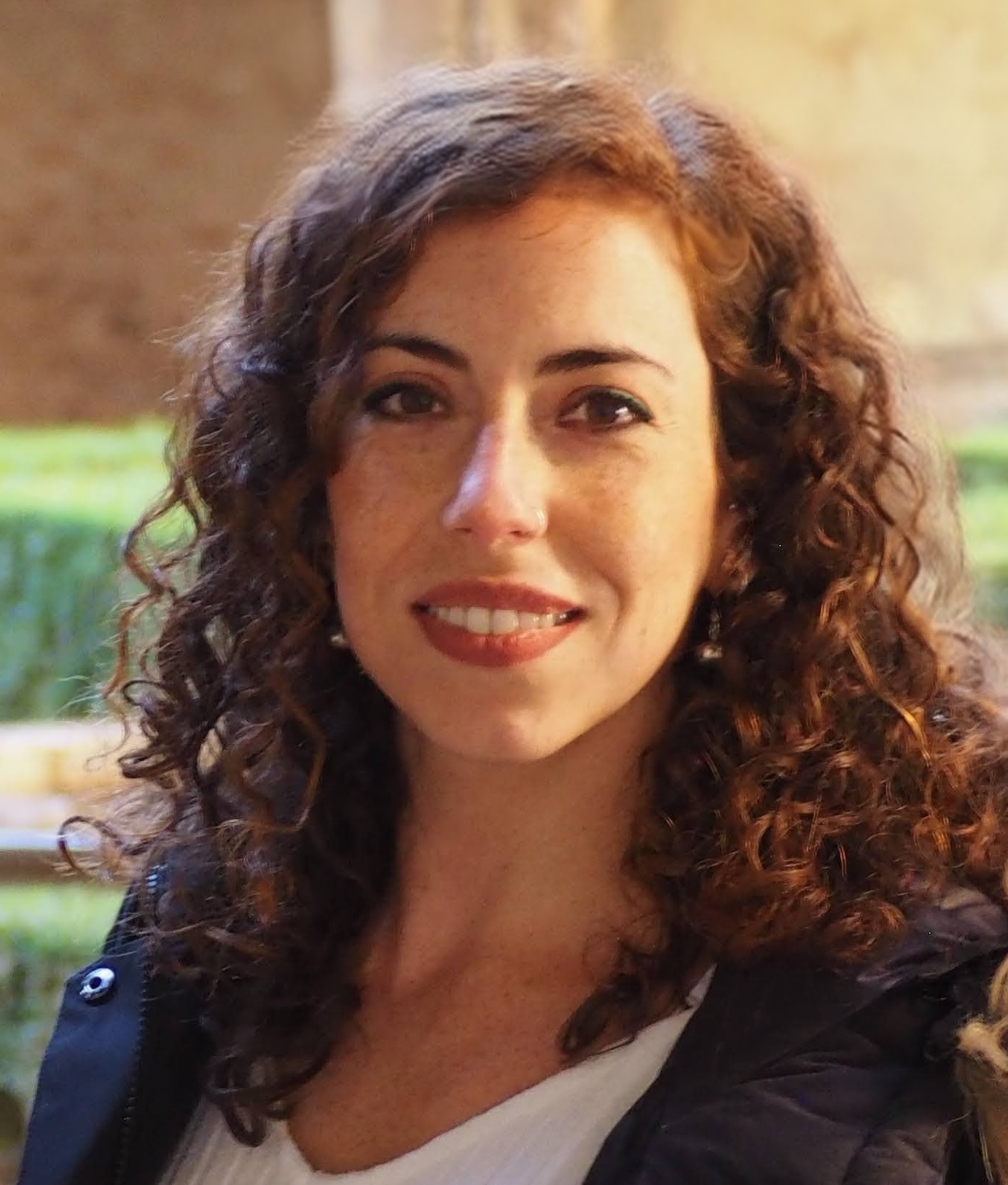}\\
  \vspace{-10pt}
\end{wrapfigure}
{\raggedright
\textbf{Lola Burgue\~{n}o}\\
	University of Malaga\\
	ITIS Software\\
	Av. Cervantes, 2, 29071, Malaga, Spain\\
	\textbf{lolaburgueno@uma.es}
\\
~
\\
}
\end{minipage}

 \noindent Lola Burgue\~{n}o is an Associate Professor at the University of Malaga (UMA). 
 Her research interests focus mainly on the fields of Software Engineering (SE) and of Model-Driven Software Engineering (MDE). She has made contributions to the application of artificial intelligence techniques to improve software development processes and tools; uncertainty management during the software design phase; model-based software testing, and the design of algorithms and tools for improving the performance of model transformations, among others. For more information, please visit \url{https://lolaburgueno.github.io/}.

\vspace{10px}

\noindent
\begin{minipage}{\columnwidth}
\begin{wrapfigure}{L}{0.3\columnwidth}
  \vspace{-20pt}
  \centering
  \includegraphics[width=0.25\columnwidth]{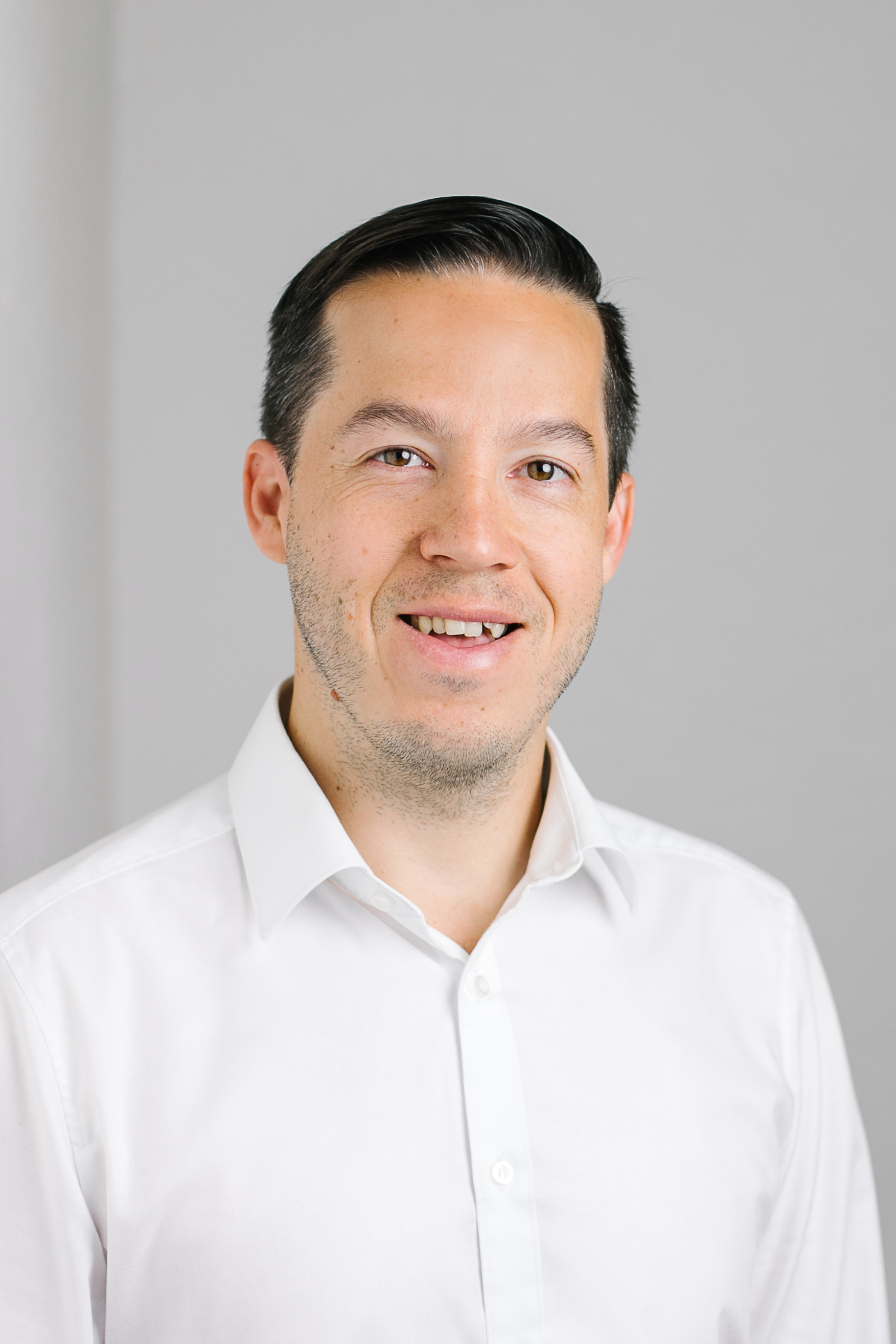}\\
  \vspace{-10pt}
\end{wrapfigure}
{\raggedright
\textbf{Dominik Bork}\\
	TU Wien \\
	Business Informatics Group\\
	Erzherzog-Johann-Platz 1, 1040, Vienna, Austria\\
	\textbf{dominik.bork@tuwien.ac.at}
\\
~
\\
~
\\
}
\end{minipage}

\noindent Dominik Bork is an Associate Professor of Business Systems Engineering at the Faculty of Informatics, Institute of Information Systems Engineering at TU Wien, where he is head of the Business Informatics Group. His research interests encompass conceptual modeling, model-driven engineering, and the development of modeling tools. A primary focus of ongoing research is on the mutual benefits of modeling and artificial intelligence. For more information, visit \url{https://www.model-engineering.info/}.\\

% BibTeX users please use one of
%\bibliographystyle{spbasic}      % basic style, author-year citations
\bibliographystyle{spmpsci}      % mathematics and physical sciences
%\bibliographystyle{spphys}       % APS-like style for physics
%\bibliography{}   % name your BibTeX data base

\bibliography{biblio}

\end{document}